\documentclass[10pt,journal,compsoc]{IEEEtran}
\ifCLASSOPTIONcompsoc
  \usepackage[nocompress]{cite}
\else
  \usepackage{cite}
\fi
\ifCLASSINFOpdf
   \usepackage[pdftex]{graphicx}
   \graphicspath{{../pdf/}{../jpeg/}}
   \DeclareGraphicsExtensions{.pdf,.jpeg,.png}
\else
\fi
\usepackage[usenames,dvipsnames,svgnames,table]{xcolor}
\usepackage{soul}
\usepackage[normalem]{ulem}
\usepackage{svg}

\usepackage{hyperref}
\hypersetup{pdftex,
   pagebackref=true,
   pdfpagemode=UseNone,
   pdftitle={\@title},
   pdfauthor={\@author},
   pdfsubject={},
   pdfkeywords={},
   pageanchor=true,
   plainpages=false,       % for problems with page referencing
   hypertexnames=false,    % for handling subfigures correctly
   bookmarks=true,         % we want bookmarks
   bookmarksnumbered=false,% include the section numbers in the list
   bookmarksopen=true,     % In the list, display highest level only
   bookmarksopenlevel=3,   % display three levels of bookmarks
   pdfpagemode=UseNone,    % show just the page
   pdfstartview=Fit,       % defail page view is the whole page at once
   pdfborder={0 0 0},      % we don't want those silly boxes for links
   breaklinks=true,        % allow breaking links
   colorlinks=false,       % don't change coloring of links
   nesting=true,
   linktocpage}

\usepackage[hang,scriptsize,sf]{subfigure}

\usepackage{tabularx}
\usepackage{colortbl}
\usepackage{tablestyles}
\newcommand{\tablerow}[8]{#1 & #2 & #3 & #4 & #5 & #6 & #7 & #8\\}
\newcommand{\chromoskeinrow}[8]{\hline\textbf{#1} & #2 & #3 & #4 & #5 & #6 & #7 & #8\\}
\newcommand{\needsserver}{$^*$}
\newcommand{\juiceboxlinking}{$^\dagger$}
\renewcommand{\to}{$\Rightarrow$\ }
\newcommand{\toot}{$\Leftrightarrow$\ }

\newcommand{\eg}{e.g.}
\newcommand{\ie}{i.e.}

\newcommand\changed[1]{\textcolor{Black}{#1}} %% Changed
\newcommand\chromoskein{\texttt{ChromoSkein}}
\newcommand\completed[1]{
  \ifnum #1=100
  \texorpdfstring{\colorbox[rgb]{0.0, 0.40784313725490196, 0.21568627450980393}{\texttt{\textcolor{White}{#1\%}}}}{#1} % 0,104,55
  \else
  \ifnum #1>89
  \texorpdfstring{\colorbox[rgb]{0.10196078431372549, 0.596078431372549, 0.3137254901960784}{\texttt{\textcolor{White}{#1\%}}}}{#1} % 26,152,80
  \else
  \ifnum #1>79
  \texorpdfstring{\colorbox[rgb]{0.4, 0.7411764705882353, 0.38823529411764707}{\texttt{\textcolor{White}{#1\%}}}}{#1} % 102,189,99
  \else
  \ifnum #1>69
  \texorpdfstring{\colorbox[rgb]{0.6509803921568628, 0.8509803921568627, 0.41568627450980394}{\texttt{\textcolor{White}{#1\%}}}}{#1} % 166,217,106
  \else
  \ifnum #1>59
  \texorpdfstring{\colorbox[rgb]{0.8509803921568627, 0.9372549019607843, 0.5450980392156862}{\texttt{\textcolor{White}{#1\%}}}}{#1} % 217,239,139
  \else
  \ifnum #1>49
  \texorpdfstring{\colorbox[rgb]{0.996078431372549, 0.8784313725490196, 0.5450980392156862}{\texttt{\textcolor{White}{#1\%}}}}{#1} % 254,224,139
  \else
  \ifnum #1>39
  \texorpdfstring{\colorbox[rgb]{0.9921568627450981, 0.6823529411764706, 0.3803921568627451}{\texttt{\textcolor{White}{#1\%}}}}{#1} % 253,174,97
  \else
  \ifnum #1>29
  \texorpdfstring{\colorbox[rgb]{0.9568627450980393, 0.42745098039215684, 0.2627450980392157}{\texttt{\textcolor{White}{#1\%}}}}{#1} % 244,109,67
  \else
  \ifnum #1>19
  \texorpdfstring{\colorbox[rgb]{0.8431372549019608, 0.18823529411764706, 0.15294117647058825}{\texttt{\textcolor{White}{#1\%}}}}{#1} % 215,48,39
  \else
  \ifnum #1>9
  \texorpdfstring{\colorbox[rgb]{0.6470588235294118, 0.0, 0.14901960784313725}{\texttt{\textcolor{White}{#1\%}}}}{#1} % 165,0,38
  \else
  \texorpdfstring{\colorbox[rgb]{0.6470588235294118, 0.0, 0.14901960784313725}{\texttt{\textcolor{White}{#1\%}}}}{#1} % 165,0,38
  \fi
  \fi
  \fi
  \fi
  \fi
  \fi
  \fi
  \fi
  \fi
  \fi
}

\begin{document}
\title{ChromoSkein: Untangling Three-Dimensional Chromatin Fiber With a Web-Based Visualization Framework}

% Mat\'{u}\v{s} Tal\v{c}\'{i}k, Filip Op\'{a}len\'{y}, Tereza Clarence, Katar\'{i}na Furmanov\'{a}, Jan By\v{s}ka, Barbora Kozl\'{i}kov\'{a}, David Kou\v{r}il
% \author{Michael~Shell,~\IEEEmembership{Member,~IEEE,}
%         John~Doe,~\IEEEmembership{Fellow,~OSA,}
%         and~Jane~Doe,~\IEEEmembership{Life~Fellow,~IEEE}% <-this % stops a space
\author{Mat\'{u}\v{s} Tal\v{c}\'{i}k,
        Filip Op\'{a}len\'{y},
        Tereza Clarence,
        Katar\'{i}na Furmanov\'{a},
        Jan By\v{s}ka,\\
        Barbora Kozl\'{i}kov\'{a},
        and~David Kou\v{r}il% <-this % stops a space
\IEEEcompsocitemizethanks{\IEEEcompsocthanksitem M. Tal\v{c}\'{i}k, F. Op\'{a}len\'{y}, K. Furmanov\'{a}, B. Kozl\'{i}kov\'{a}, D. Kou\v{r}il are with Masaryk University, Brno, Czechia.
\IEEEcompsocthanksitem J. Byška is with Masaryk University, Brno, Czechia and University of Bergen, Bergen, Norway.
\IEEEcompsocthanksitem T. Clarence is with Icahn School of Medicine at Mt Sinai, New York, NY, USA.
\IEEEcompsocthanksitem Correspondence to D. Kou\v{r}il (dvdkouril@mail.muni.cz) and T. Clarence (tereza.clarence@mssm.edu).}% <-this % stops an unwanted space
\thanks{Manuscript version: October 11, 2022.}}

\markboth{}%
{Shell \MakeLowercase{\textit{et al.}}: Bare Demo of IEEEtran.cls for Computer Society Journals}



\IEEEtitleabstractindextext{%
\begin{abstract}
\changed{We present ChromoSkein, a web-based tool for visualizing three-dimensional chromatin models. The spatial organization of chromatin is essential to its function. Experimental methods, namely Hi-C, reveal the spatial conformation but output a 2D matrix representation. Biologists leverage simulation to bring this information back to 3D, assembling a 3D chromatin shape prediction using the 2D matrices as constraints.}
\changed{Our overview of existing chromatin visualization software shows that the available tools limit the utility of 3D through ineffective shading and a lack of advanced interactions. We designed ChromoSkein to encourage analytical work directly with the 3D representation. Our tool features a 3D view that supports understanding the shape of the highly tangled chromatin fiber and the spatial relationships of its parts.}
\changed{Users can explore and filter the 3D model using two interactions. First, they can manage occlusion both by toggling the visibility of semantic parts and by adding cutting planes. Second, they can segment the model through the creation of custom selections.}
\changed{To complement the 3D view, we link the spatial representation with non-spatial genomic data, such as 2D Hi-C maps and 1D genomic signals.
% Put together, our tool allows analysis of chromatin’s spatial conformation beyond what is possible using purely non-spatial visualization.
We demonstrate the utility of ChromoSkein in two exemplary use cases that examine functional genomic loci in the spatial context of chromosomes and the whole genome.}
\end{abstract}

}

\maketitle

\IEEEdisplaynontitleabstractindextext

% \fi
\IEEEpeerreviewmaketitle

\IEEEraisesectionheading{\section{Introduction}\label{sec:introduction}}

% for IEEE Computer Society journal papers produced under \LaTeX\ using
% IEEEtran.cls version 1.8b and later.
% % You must have at least 2 lines in the paragraph with the drop letter
% % (should never be an issue)
% I wish you the best of success.

\IEEEPARstart{O}{rganization}
of a genome in three-dimensional (3D) space significantly impacts its function. In eukaryotic cells, the DNA is packed into a micron-sized nucleus with the help of proteins, which together build a so-called \emph{chromatin fiber}.
Chromatin represents a unique combination of two data characteristics. First, the underlying DNA molecule is a \textbf{linear} nucleotide sequence. Second, the fiber folds into a densely packed \textbf{3D} shape.
%% , resulting in a dense packing inside a cell nucleus.

The specific spatial configuration of chromatin in a nucleus is not yet fully determined. Thus, it is an object of intense study in biology. Over the years, several methods for revealing the organization of chromatin have been developed. Most prominent is the \emph{chromosome conformation capture (3C)} suite of experiments~\cite{han:3c:2018}.
These methods cross-link together segments of the DNA that are in physical proximity.
Aggregating the resulting interactions across many cells gives a measure of how likely two genomic parts are mutually interacting.
The \emph{Hi-C} method~\cite{lieberman-aiden:2009:hi-c}, a variant of 3C, resolves all-to-all interactions between same-sized DNA segments.
The output of this method is usually in the form of a contact matrix, see~\autoref{fig:whole-pipeline}.
\changed{The matrix values can then be used to assemble a prediction of chromatin's conformation in 3D.}

% MAYBEDELETE
% \changed{Existing chromatin visualization tools predominantly showcase and leverage the linearity aspect.}
% \changed{So-called \emph{genomic browsers} map different characteristics, called \emph{genomic signals}, onto 1D genomic coordinates and allow users to interactively scroll along the sequence.}
% \changed{Similarly, \emph{Hi-C matrix viewers} lay out data along the linear sequence.}
%
%% 3D aspect
\changed{While a number of tools feature a 3D view, the 3D aspect is underutilized in genomic visualization.}
We hypothesize that the lack of tools supporting 3D interaction is caused mainly by two reasons. First, the Hi-C experiment outputs 2D contact maps. Therefore, most tools naturally work with the 2D representation.
Second, 3D visualization inherently presents issues like occlusion that make it harder to work with 3D data on a two-dimensional display.
%% DK: potentially could delete
Nusrat et al.~\cite{nusrat-2019} further remark that visualizing 3D spatial chromatin data suffers from the fact that it only shows a prediction and omits information about the resolved structure's ambiguity.

Despite these claims, certain theories can only be examined in the 3D context.
For example, positioning of a gene in a chromosome and its spatial distance to functional landmarks, such as promoters and enhancers, has a direct impact on gene regulation~\cite{dekker:hierarchy:2013}. 
\changed{Certain data characteristics---\eg, density or mutual spatial orientation---are more intuitively obtained from the 3D representation.}
The examined structure, \ie, chromatin fiber, is by definition a spatial object and, according to recent biological research, it makes sense to examine it in its natural context~\cite{chi:2022:3d-chromosome-architecture}.
\changed{We found out that many tools use 3D chromatin models merely for illustration while interaction with the 3D data is severely limited.}

\changed{To address this gap in available chromatin visualization tools, we set out to design and develop a novel tool that places the 3D representation at the forefront.}
\changed{We focus primarily on two facets: a) performant and high-fidelity visual representations that highlight the spatiality of the underlying data; and b) interaction allowing direct exploration of the 3D chromatin model.}
\changed{We deliver our visualization solution for the web environment to promote easier adoption and collaboration.}

In summary, we present the following key contributions:
%% \begin{itemize}[topsep=0pt,itemsep=0ex,partopsep=1ex,parsep=0ex]
\begin{itemize}
\item \changed{An extensive overview of existing genomic tools featuring 3D visualization, 
% leading to definition of requirements for a 3D-centric chromatin system.
reviewing available visual representations, interactions, and analytical features.} %% DK: is 'interactions' and 'analytical features' different?
% \item \dk{Description of typical domain tasks performed on 3D chromatin data leading into a formulation of requirements for a visualization system.}
\item \changed{Design and prototypical implementation of a 3D-centric visualization tool for chromatin data, called \chromoskein\footnote{Available at \texttt{\url{https://github.com/chromoskein/chromoskein}}}.} %% DK: 'web-based' somewhere?
\item \changed{Discussion and demonstration of ways of linking the 3D representation with conventional genomic 1D and 2D visualizations.}
%% \item \dk{A prototypical implementation of presented concepts in a novel tool called \emph{Chromoskein}.}
\item \changed{Two exemplary use case scenarios demonstrating the utility of 3D representation in domain-specific analytical tasks.}
\end{itemize}

\section{Background \& Motivation}
\label{sec:background}
Our work is motivated by a specific biological application domain. We, therefore, start with a brief introduction to chromatin conformation research
%% , which is essential for understanding the terminology used throughout the paper.
%% Further, we
and
outline the high-level motivation for developing a novel tool.

\subsection{Background}
%% The linear structure of the DNA is defined by a sequence of nucleotides, each carrying one of the four nucleobases: adenine, cytosine, guanine, or thymine.
While methods for resolving the linear DNA sequence are well established, the acquisition of DNA's three-dimensional organization is still challenging. It is known that the double helix is further organized in nucleosomes, which in turn form a chromatin fiber packaged into chromosomes in the cell nucleus. 

The organization of chromatin in a cell is the subject of several experimental methods, referred to as chromosome conformation capture (3C) techniques~\cite{han:3c:2018}. These methods can indicate the spatial organization of chromatin. 
In general, 3C techniques are based on measurements of the frequency of interactions between fragments of DNA.
%% parts of a chromatin fiber---fragments of DNA.
Earlier methods could yield results only for a few of these fragments.
%% \annot{hi-c}
However, high-throughput variations of 3C methods, such as Hi-C~\cite{lieberman-aiden:2009:hi-c, belton:20122:hic}, enable biologists to infer the number of interactions between all equally-sized regions of the whole chromatin fiber.
%% This way, they can determine the spatial conformation of the whole fiber, but
The trade-off is a very limited precision.
%% \annot{bins}
Depending on the experiment's parameters, the size of these regions is in the order of thousands (kilobases,~kb) to millions of nucleotides (megabases, Mb). One such group of nucleotides is then referred to as a \emph{bin}.
The output of a Hi-C assay is a 2D \emph{contact frequencies matrix}, sometimes also called \emph{interaction frequencies matrix}, that assigns each pair of bins a number proportional to the number of interactions observed between these two bins. %% DK plural or singular in matrix adjective?

%% \annot{using the matrices}
Further analysis of Hi-C matrix data reveals multi-scale organization of genomes.
The organization manifests in the Hi-C map as specific patterns, \eg, a chessboard-like pattern signifying A/B compartments~\cite{lieberman-aiden:2009:hi-c}, or point peaks in distant sequential regions indicating topologically associating domains (TADs), important in gene expression~\cite{dixon:2012:topological}. %% DK: these two references taken from Stevens 2017

\begin{figure}
%% DK: (a), (b) captions are randomly centered: probably there is a better way to do this in later revision

  \centering
  \includegraphics[width=\linewidth]{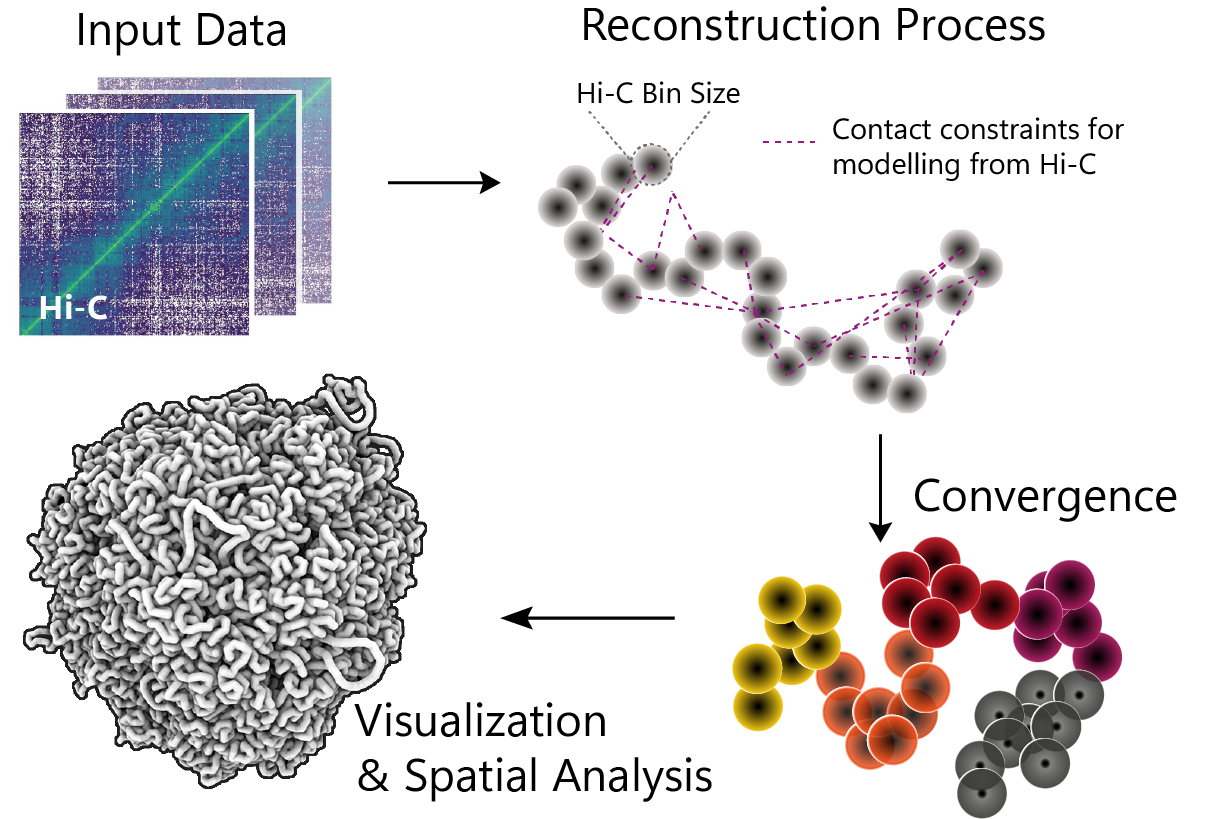}
\vspace{-4mm}
\caption{Overview of reconstruction process: first, contact frequency data are measured using Hi-C experiments. These are used as a constraint in modelling algorithms. After that, chromatin can be studied in 3D.
}
  \label{fig:whole-pipeline}
\vspace{-2mm}
\end{figure}

% \todo{Merge this part with stuff above (moved here from data).} 
The contact frequency maps coming from the Hi-C experiment imply a genome's spatial configuration.
%% but hide certain spatial aspects.
%
Computational biologists thus infer the 3D structure using probabilistic algorithms, polymer simulations, or various other methods~\cite{Oluwadare:2019:rec3D}. The Hi-C matrix values serve as constraints for the computation\changed{, as illustrated in~\autoref{fig:whole-pipeline}}. These methods generate a three-dimensional \emph{model prediction}, where every bin is assigned a position in 3D space
%% , assuming the most probable position of the group of nucleotides it represents.
Examples of tools for genome structure prediction are \texttt{LorDG}~\cite{trieu-cheng:2016:lordg} or \texttt{3DMax}~\cite{oluwadare:2018:3dmax}.

\subsection{Biological Motivation}
The three-dimensional chromatin structure can be used to analyze and confirm hypotheses.
%% posed by the experts.
For example, Stevens et al.~\cite{tan:2018:3d-diploid}, who first reconstructed whole-genome models using single-cell Hi-C from haploid mouse cells, examined the 3D model to confirm known spatial features. 
They found that chromosomes occupy specific areas---territories---in the nucleus. They did so by isolating individual chromosomes and de-emphasizing the rest to indicate the chromosomes' nuclear position.

Similarly, Tan et al.~\cite{tan:2018:3d-diploid} reconstructed whole genome chromatin models but from diploid human cells. They also look at chromosome territories.
The shape and mutual arrangement of chromosomes are evident from the 3D model, such as how the chromosomes intermingle.
%% Another interesting spatial feature is the distance from the nuclear center.
Tan et al. also show how gene-rich chromosomes occupy space closer to the nuclear center while gene-poor chromosomes lie on its periphery.
With the 20kb resolution, the 3D model exhibits a hierarchical, fractal organization of chromatin, with regions of high spatial clustering and segregation.
%% These levels can be examined by zooming and filtering.
Thanks to Tan et al.'s novel method for capturing conformation of diploid cells, the resulting structures showcase differences in the shape of active and inactive X chromosomes. While this can be quantified by, for example, a radius of gyration, this feature can be intuitively identified during the visual exploration of the 3D shape.

%% \annot{motivation}
The development of \chromoskein~is motivated by these recent investigations of chromatin conformation. Spatial representations clearly play an important role in several stages of the analytical process.
We see an opportunity for intuitive visualization and interaction methods to expedite the analysis of 3D chromatin data.
%% In this area, most of the figures are generated using custom scripting. However, the ability to explore, analyze, and prepare data for presentation in the same context, would help in formulating and confirming new hypotheses. %could be beneficial.

\subsection{\changed{Why Focus on 3D?}}
%% DK: calculating human and mouse genome length to (theoretical) bins:
%% mouse
%% total bases: 2,728,222,451 from https://www.ncbi.nlm.nih.gov/grc/mouse/data
%% @15kb
%% 2728222451 / 15000 = 181881.496733 -> 181 881
%% human
%% genome size: 3,117,275,501 bp from https://en.wikipedia.org/wiki/Human_genome
%% @15kb
%% 3117275501 / 15000 = 207818.366733 -> 207 818

\changed{Bioinformaticians trained in Hi-C analysis
%% are used to looking at contact matrices and
can imply insights from the contact matrix representation.
Nevertheless, they are still limited by their cognitive abilities. Hi-C maps are an abstract representation that captures many relationships between genomic regions.
%% Putting all these constraints together is beyond one's abilities.
3D structure prediction allows generating a spatial model that satisfies all the constraints and gives one possible conformation in space. This relieves scientists of having to reconstruct a mental image of the spatial conformation.}
%% and instead allows them to focus on generating hypotheses based on the specific instance.}

\changed{Additionally, certain characteristics of a spatial structure cannot be inferred purely from the 2D representation. For example, biologists might be interested in density of functional landmarks (\eg, genes) in a certain nucleus compartment. This metric can only be determined from the modeled 3D structures, either visually or by radius of gyration.
Similarly, mutual orientation of elements, \eg, chromosomes, is a relationship indirectly captured in the Hi-C map but observable only once the structure is visualized in 3D.}

\begin{figure*}[t]
 \centering % avoid the use of \begin{center}...\end{center} and use \centering instead (more compact)
 \includegraphics[width=\textwidth]{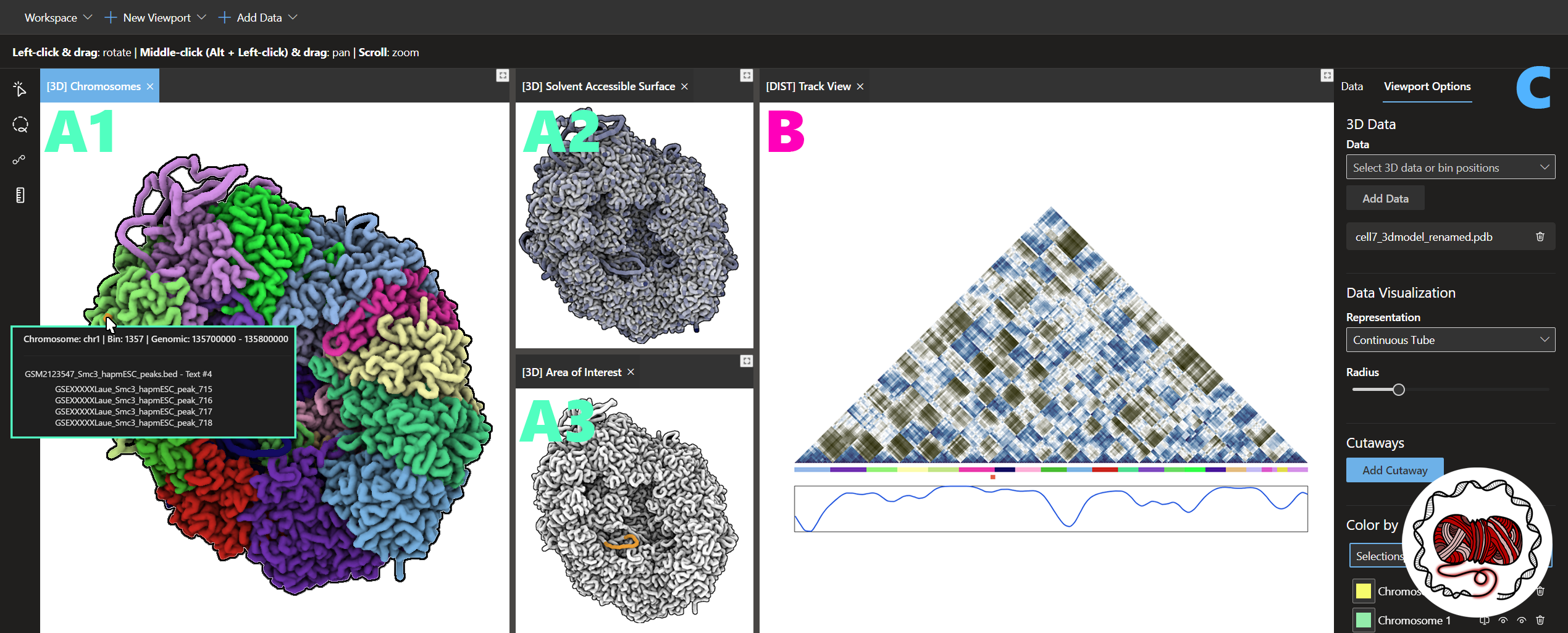}
 \vspace{-5pt}
 \caption{Overview of \chromoskein. A) 3D viewports. A1) Colored by imported chromosomes. Shown with tooltip on the hovered bin. A2) Colored by solvent accessible surface area  (SASA) computed by \chromoskein. A3) With hidden chromosome coloring but with the visible user-made selection of interest. B) Track view with distance map of the 3D structure shown along with chromosome selections track, user-made selection track, and 1D SASA track. C) Configuration panel.} %Overview of \chromoskein. A) Main 3D viewport, showcasing highlighted selections. A2) On-hover tooltip of the bin. B) Distance viewport, with selections shared with the 3D viewport in part A. C) Secondary 3D viewport with the same dataset as in A but with hidden non-selected parts and mapped distance from centromeres. D) Tool panels: the left panel for tool selection, the top panel for tool configuration, and the right panel for configuration of the currently selected viewport. \dk{TODO: should be extended into a full-length figure, also indicate inputs (PDB, 1d stuff etc.)}
 \label{fig:overview}
  \vspace{-5pt}
\end{figure*}

%% \annot{space argument}
\changed{Finally, Hi-C maps are often quite extensive, occupying gigabytes in storage.
At a common 15kb resolution, where one bin corresponds to a sequence of 15 thousand basepairs, a human genome results in approx. 207 thousand bins. Storing the information about observed interactions for all possible pairs of bins thus results in ${207\,000} ^ 2$ numbers.
As the Hi-C map is symmetric, some viewers only show half of it, \ie, triangular Hi-C map (see \autoref{fig:overview}B), resulting in ${N}^2/2$ numbers to store. However, the asymptotic complexity $O(n^2)$ remains the same.}
%\changed{Finally, Hi-C maps are often quite extensive, occupying gigabytes in storage. At a common 15kb resolution---\ie, one bin corresponds to a sequence of 15 thousand basepairs---a human genome theoretically results in approx. 207 thousand bins.
%Hi-C map contains values of how likely each pair of bins is interacting. In the exemplary case of the human genome, this equals to storing ${207\,000} ^ 2$ numbers\footnote{\changed{As the Hi-C map is symmetric, some viewers only show half of it, \ie, triangular Hi-C map, resulting in ${N}^2/2$ numbers to store; the asymptotic complexity, however, remains the same.}}.}
\changed{The 3D representation, on the other hand, requires only a list of bin positions, reducing the space complexity from $O(n^2)$ to $O(n)$. The 3D model thus serves as a more efficient form for storage and transfer.}

%% DK: make a note here about how you can get the distance map from 3D and how that's pretty much the same as hi-c map?

\changed{Naturally, the 3D representation has its disadvantages. As Nusrat et al.~\cite{nusrat-2019} note, 3D chromatin models represent only a prediction that satisfies the constraints defined by the Hi-C map.
This uncertainty is, however, understood and accounted for by biologists investigating 3D chromatin.
3D visualization also inherently presents issues related to showing 3D phenomena on a 2D display, such as occlusion. Occlusion management, however, has been a topic of research both in general computer graphics and in visualization specifically and we now have at our disposal a number of techniques to deal with this issue.}
\changed{Overall, we do not argue that a 3D representation should completely replace the 2D matrix representation. We do, however, believe that it can serve as a valuable addition and that existing tools underutilize this representation.}

\section{Related Work}
\label{sec:rw}
While building on general molecular visualization, tools visualizing genomic data have developed in their own strand of research. Here we review visualization's role in the investigation of genomes' spatial organization and survey existing visualization tools.
\changed{\autoref{tab:RW} summarizes the reviewed tools and highlights available and missing functionality.}

%% vvvvvvvvv RELATED WORK TABLE HERE vvvvvvvvv
\newcolumntype{x}{>{\centering\arraybackslash}X}
\newcolumntype{N}{>{\hsize=1.0\hsize}X}
\newcolumntype{r}{>{\hsize=1.2\hsize}x}
\newcolumntype{a}{>{\hsize=1.0\hsize}x}
\newcolumntype{b}{>{\hsize=1.1\hsize}x}
\newcolumntype{m}{>{\hsize=0.8\hsize}x}
\newcolumntype{o}{>{\hsize=0.9\hsize}x}
\newcolumntype{s}{>{\hsize=0.5\hsize}x}
\begin{table*}[ht!]
\tablestyle[sansbold]
\footnotesize
\begin{tabularx}{\linewidth}{ N | rxmboss}
\arrayrulecolor{lightgray}
\theadstart
    \thead Tool &
    \thead Hi-C Representations &
    \thead Annotations &
    \thead Selections &
    \thead View linking &
    \thead Occlusion management &
    \thead Depth-cueing shader &
    \thead Web-based \\
\tbody

%  \tablerow{Juicebox}{2D Heatmap}{1D Tracks, 2D Glyphs}{Chromosome, Square}{2D \to 2D\juiceboxlinking}{---}{Not applicable}{No}
%  \tablerow{Juicebox.js}{2D Heatmap}{1D Tracks, 2D Glyphs}{Chromosome, Square}{2D \to 2D\juiceboxlinking}{---}{Not applicable}{Yes}
%  \tablerow{HiGlass}{2D Heatmap}{1D Tracks, 2D Glyphs}{Chromosome, Rectangle}{2D \toot 2D}{---}{Not applicable}{Yes\needsserver}
%  \tablerow{3DIV}{2D Heatmap, Arcplot}{1D tracks}{}{}{---}{Not applicable}{Yes\needsserver}
%  \hline

\tablerow{Genome3D \cite{asbury:2010:genome3d}}{Tube, Spheres, Glyphs}{Coloring}{Selection Gizmo}{---}{---}{Yes}{No}
\tablerow{3DGB \cite{butyaev:2015:3dgb}}{Tube}{Coloring}{Cube, Linear}{1D \toot 3D}{---}{No}{Yes\needsserver}
\tablerow{GMOL \cite{nowotny:2016:gmol}}{Ball and Stick}{Coloring, Text labels, Interaction lines}{---}{---}{---}{No}{No}
\tablerow{3Disease \cite{li:2016:3disease}}{Tube, 2D Heatmap}{Coloring}{---}{?}{?}{No}{Yes\needsserver}
\tablerow{TADkit \cite{tadkit-github}}{Tube, 2D Heatmap}{Coloring}{---}{1D, 2D \to 3D}{---}{No}{Yes}
\tablerow{HiC-3DViewer~\cite{djekidel:2017:hic-3dviewer}}{Tube}{Coloring, Interaction lines}{---}{1D, 2D \to 3D}{---}{No}{Yes\needsserver}
\tablerow{Delta \cite{tang:2017:delta}}{Tube, Spheres}{Coloring, Labels, Interaction lines}{Point, Linear, Loop}{1D \toot 3D}{---}{No}{Yes\needsserver}
\tablerow{GenomeFlow~\cite{trieu:2019:genomeflow}}{Tube}{Coloring, Labels, Line thickness}{Drag rectangle}{---}{Hide regions}{No}{No}
\tablerow{CSynth \cite{todd:2020:csynth}}{Tube}{Coloring, Labels}{Point, Gene}{2D \toot 3D}{---}{No}{Yes}
\tablerow{Nucleome Browser \cite{zhu:2022:nucleome-browser}}{Line, Tube, Sphere, Billboard crosses}{Coloring}{Point, Chromosome}{1D, 2D, 3D \toot 3D}{Hide regions}{No}{Yes\needsserver}
\tablerow{SpaceWalk \cite{spacewalk2022}}{Line, Tube, Sphere, Point cloud}{Coloring, Labels, Glyphs}{Point}{1D \to 3D}{---}{No}{Yes}
\tablerow{WashU Epigenome Browser \cite{wang:2019:washu}}{Line, Tube, Sphere, Billboard crosses}{Coloring, Labels, Glyps, Arrows}{Point}{1D, 2D \toot 3D}{Hide uniparental chromosomes, Line opacity}{No}{Yes\needsserver}
\chromoskeinrow{Chromoskein}{Line, Tube, Sphere}{Coloring, Glyphs}{Point, Linear, Sphere}{2D, 3D \toot 3D}{Cutting planes, Hide regions}{Yes}{Yes}
\end{tabularx}
\label{tab:style:sansbold}
\vspace{0.5em}

\justifying
\noindent\needsserver\ Needs a server for computations or data management.

\noindent\juiceboxlinking\ Restricted to two views.

\noindent ?\ Unknown due to unavailability of the tool and/or missing information in the accompanying publication.
\vspace{1em}
\caption{Comparison of 3D visualization genomic tools for three-dimensional datasets.}
\vspace{-7mm}
\label{tab:RW}
\end{table*}

%% \annot{surveys and overviews}
Marti-Renom and Mirny~\cite{marti-renom:2011:bridging-the-gap} discuss the challenges and approaches in resolving the biological structure spanning several magnitudes of scales with available experimental and imaging technology.
Goodstadt and Marti-Renom~\cite{Goodstadt:2019:communicating-genome-arch} further survey means of visualizing existing genomic data types and list available software tools.
Waldisp\"{u}hl~et~al.~\cite{waldispuehl-2018} provide additional insight into the technical challenges related to visualizing 3D genomes.

%% \annot{2D}
Initially, chromatin structure was examined purely using {Hi-C} matrices, \ie, in 2D representation. \texttt{Juicebox}~\cite{durand:2016:juicebox} first enabled interactive zooming in the large-scale space of contact frequencies. Its authors leverage a tiling approach inspired by web map services, such as Google Maps. \texttt{HiGlass}~\cite{kerpedjiev:2018:higlass} further expanded on the idea of zoomable Hi-C maps by implementing additional capabilities, \eg, several linked views and juxtaposing genomic (mostly 1D) data to the Hi-C map.
Other visualization types complementing the Hi-C matrices are also common, such as the arc-plot variations used by \texttt{3DIV}~\cite{Yang:2017:3DIV}.

%% \annot{in VIS field}
Most publications on visualizing genomes come from bioinformatics research and therefore largely focus on the applications. Nusrat et al.~\cite{nusrat-2019} provide a comprehensive survey of the topic from the visualization research perspective.
Recently, L'Yi et al.~\cite{lyi-2022-gosling} introduced \texttt{Gosling}, a framework for genomic data visualization that mostly focuses on 2D plots. They leverage a grammar-based approach popularized in the visualization community by Vega-Lite~\cite{satyanarayan:2017:vega-lite}.

%% \annot{3D: abusing molecular tools}
Working with \emph{three-dimensional} genomic data presents several additional challenges. Goodstadt and Marti-Renom~\cite{goodstadt-2017} debate the challenges of visualizing 3D genomes directly.
Many domain experts employ existing molecular graphics tools, \eg, \texttt{PyMol}~\cite{PyMOL} or \texttt{ChimeraX}~\cite{pettersen:2021:chimerax}, to perform the analysis of 3D chromatin.
These tools often have a decades-long history and over time have developed into colossal suites tailored mostly for the analysis of smaller molecules, such as proteins. Their applicability to 3D genomic data is limited due to a high learning curve and sometimes technical limitations on larger datasets.
Furthermore, including other views (\eg, Hi-C matrix) requires external software or extensive scripting.

%% \annot{first 3D genome vis: desktop}
One of the first tools tailored to the exploration of the genome in its spatial context is \texttt{Genome3D}~\cite{asbury:2010:genome3d}.
The tool allows switching between three levels of scale that correspond to the inherent hierarchy of the genome.
\texttt{GMOL}~\cite{nowotny:2016:gmol} expanded on \texttt{Genome3D} and increased the number of explorable levels to six. Interaction is mostly done through command-line input.
Both of these tools were created as desktop programs, limiting wider adoption in research.

%% \annot{web}
Software published as a web application, on the other hand, is instantly available. In genomic research, tools like the \texttt{UCSC Genome Browser}~\cite{kent:2002:ucsc} benefited from the decision to target the web platform.
Consequently, many tools for exploring 3D genomic data were developed for the web to lower the adoption threshold.
\texttt{3DGB}~\cite{butyaev:2015:3dgb} devises a solution for storage, querying, and mesh-based rendering of 3D genomic data.
%% \dk{Limits: ...}
%
Li et al.~\cite{li:2016:3disease} use \texttt{3Disease} to analyze spatial chromatin rearrangement in cancer and developmental diseases. They employ a plotting library to visualize small genomic sub-parts: TADs.
%% \dk{Limits: ...}
%
TADs can be more closely examined in \texttt{TADkit}~\cite{tadkit-github}.
%% \dk{Limits: ...}
TADkit is a WebGL-based viewer for the analysis of TADs utilizing TADbit~\cite{serra-2017-tadbit} library developed by the same team.
%% \dk{Limits: ...}
%
\texttt{CSynth}~\cite{todd:2020:csynth} combines 3D structure modeling from contact frequencies with an interactive visualization of the result, allowing human-in-the-loop workflow.
\texttt{Delta}~\cite{tang:2017:delta} features a view able to show 3D conformation of small part of the whole genome.

Many of the mentioned tools begin with a form requiring to specify parts of the dataset, \eg, genomic coordinates range, to visualize and therefore inhibit holistic analysis of both local parts as well as global context.
%
%% \annot{actual contenders}
Furthermore, while the above-listed tools feature 3D visualization in some form, this aspect is often underutilized and used only for illustration, while the actual analysis is done in other views, \eg, 2D Hi-C contact maps, or 1D feature tracks.
We identified four tools where 3D views play a larger role within the analytical workflow and allow performing tasks directly in the 3D representation.

\texttt{GenomeFlow}~\cite{trieu:2019:genomeflow}, a continuation of \texttt{GMOL}, offers functionality for modeling and analysis of 3D genomic data.
Users can examine predicted 3D structures and augment them by loading additional data, \eg, a list of chromatin loops or gene annotations, which are then overlaid over the 3D model.
Interaction with the 3D model itself is, however, limited. GenomeFlow does not allow visibility management and bins cannot be selected from the 3D view.
Trieu et al.~\cite{trieu:2019:genomeflow} also do not mention the ability of linked views and how the interaction between the 2D Hi-C maps and 3D predicted structure is coordinated.
The tool is implemented as a desktop application and has not been maintained recently which, in our opinion, limits its utility for recent datasets. 

Similarly, \texttt{HiC-3DViewer}~\cite{djekidel:2017:hic-3dviewer} combines modeling and visualization albeit on web using client-server architecture.
To work with the 3D structure, it offers mapping of genomic signals onto the 3D model.
The 3D view is complemented with a pop-up Hi-C map and a 1D track panel. Selection in the Hi-C map is reflected in the 3D view, but selection in the opposite direction is not possible.
The rendering uses flat colors without shading, limiting the perception of the chromatin's overall shape. The tool does not include cutaways or filtering, leading to high visual complexity. %visual clutter being unaddressed in large and dense datasets.

%% (vis.nucleome.org)
\texttt{Nucleome Browser}~\cite{zhu:2022:nucleome-browser} developed by the 4D Nucleosome consortium is the third relevant tool we identified.
It combines linked views offering all possible modalities: 1D for genomic signal tracks, 2D for Hi-C contact maps, and 3D for predicted 3D chromatin structure.
The overall functionality and interaction inside the 3D view are rudimentary.
The browser allows switching between global and local views but only at three granularities.
%% (chromosome, region, highlight)
Arbitrary selections are only unidirectional: users can select genomic regions in the 1D or 2D views which are reflected in the 3D, but not the other way around. Only a whole chromosome or a single bin can be selected from the 3D view.  

\changed{\texttt{WashU Epigenome Browser}~\cite{wang:2019:washu} is a genomic browser combining and linking different views. Its 3D view implements bidirectional linking in a limited way, allowing only single-point selections.
This tool, however, offers a large number of options for annotating the
3D structure. Besides the prevalent coloring by
numerical values, it can label the 3D structure with
text and glyphs.}

% %% TC: "Figure3: Schematic of ChromoSkein design and user interface"

%% \annot{our key improvements}
%% Problem: shitty 3d viewers
Currently, most web-based 3D genome visualizations use \mbox{WebGL} or libraries built on it, \eg, three.js. WebGL lacks in capabilities compared to graphics APIs for native desktops. This limits both the data sizes web viewers can render interactively and the visual fidelity.
%% Our Solution:
%We address this issue by basing our 3D rendering on the emerging WebGPU standard that offers low-level access to GPU on web.
%% Problem: no occlusion management
Furthermore, visualizing large chromatin datasets opens up an issue with occlusion, which inhibits a proper exploration of the 3D dataset. None of the reviewed existing tools include any techniques for occlusion management to allow peeking into a dense 3D dataset, apart from a few tools that implement hiding selected regions.
%% Our Solution
%We thus include several basic occlusion management strategies that reveal features hidden by the outermost layer of chromatin fiber.
%% Problem: no interaction with the 3D
Finally, a lack of interaction options for exploring and manipulating the 3D genome model is prominent across all the surveyed tools.
For the most part, the linking functionality is implemented in the 1D-to-3D or 2D-to-3D direction
%% (\ie, selecting a bin in a 2D Hi-C map highlights the bin in the 3D window),
but the other way around
%% (\ie, selecting bin in the 3D view and have the selection mirrored in the 1D and 2D views)
is often limited. Users can usually select only singular bins, and have to turn back to 1D and 2D views for advanced selections. This hinders the exploration and prevents reasoning about the observed chromatin structure.

\section{Requirements}
\changed{We base our requirements on discussions with a domain scientist during the initial phases of our collaboration. Furthermore, we consider our investigation of related tools and the exposed feature set limits detailed in the previous section.}
\changed{As a result, we define the following requirements for a novel chromatin visualization tool:}

\begin{itemize}
\item \changed{\textbf{Efficient 3D rendering:} The tool should be able to handle large datasets with sizes in the magnitude of hundreds of thousands of bins and render them in real-time with interactive framerates (30+ FPS).}
\item \changed{\textbf{Support shape comprehension:} The visual representations should highlight the spatiality of data and help users to gain an understanding of the shape and size of the model.}
\item \changed{\textbf{Visibility management:} Chromatin models tend to be dense and highly occluded. Biologists need be able to strategically highlight salient parts while removing the occlusion that prevents localizing them. Only the least amount of information should be removed to preserve context.}
\item \changed{\textbf{Bi-directional linking:} Users should be able to select genomic loci in the 3D view or the other (1D and 2D) representations. These selections should be mirrored in the remaining views as they are all better suited to different tasks. After localizing a significant part in the 3D view, it may be desired to look at multiple correlating data in several tracks of 1D views. }
\end{itemize}

\changed{The requirements served as constraints and guiding principles for the design of our novel tool, which we describe in the following section.}

% \section{\dk{2D vs 3D Representation (DK \completed{50})}}
% \input{sections/2d-vs-3d.tex}

\section{3D-Centric Design of ChromoSkein}
% \input{sections/method.tex}
% \dk{Considering the lack of tools implementing a fully functional 3D chromatin viewer, we set out to develop our own visualization tool able to support analytical tasks in the 3D representation.}
%% \changed{Considering the lack of tools implementing an advanced 3D chromatin viewer, we set out to develop our visualization tool offering a combination of all functionality outlined in \autoref{tab:RW}.}
\changed{In this section, we describe our design choices and detail the implementation of a 3D-centric chromatin visualization tool, shown in~\autoref{fig:overview}, which we call \chromoskein.}
\changed{We describe how we turn the 3D data into visual representations and how we render and shade the visual marks. Afterward, we discuss interactions with the 3D visualization.}

\subsection{Visual Representations} %% how we turn data into visual marks
%% \annot{moved from bg}
%% 3D chromatin structure produced by the computational methods from Hi-C maps is generally represented as a set of bins, each with its positional attribute.
%% The set is ordered, \ie, the order of bins corresponds to the order of individual segments in the DNA sequence.
%% Optionally, information about segmentation into chromosomes is provided as well.
%% Unfortunately, there is no single established standard format for spatial chromatin data yet.
%% However, two commonly used formats are PDB and CSV. PDB, originally designed to describe protein structures, is commonly adapted for reconstructed chromatin structures.
%% The segmentation of chromatin fiber into chromosomes is then encoded in the PDB file's connectivity records---each fully connected component represents one chromosome.
%% The second format is CSV with columns defining X, Y, and Z coordinates and optionally also a chromosomal identifier.

\begin{figure}[t!]
 \centering 
 \includegraphics[width=0.95\columnwidth]{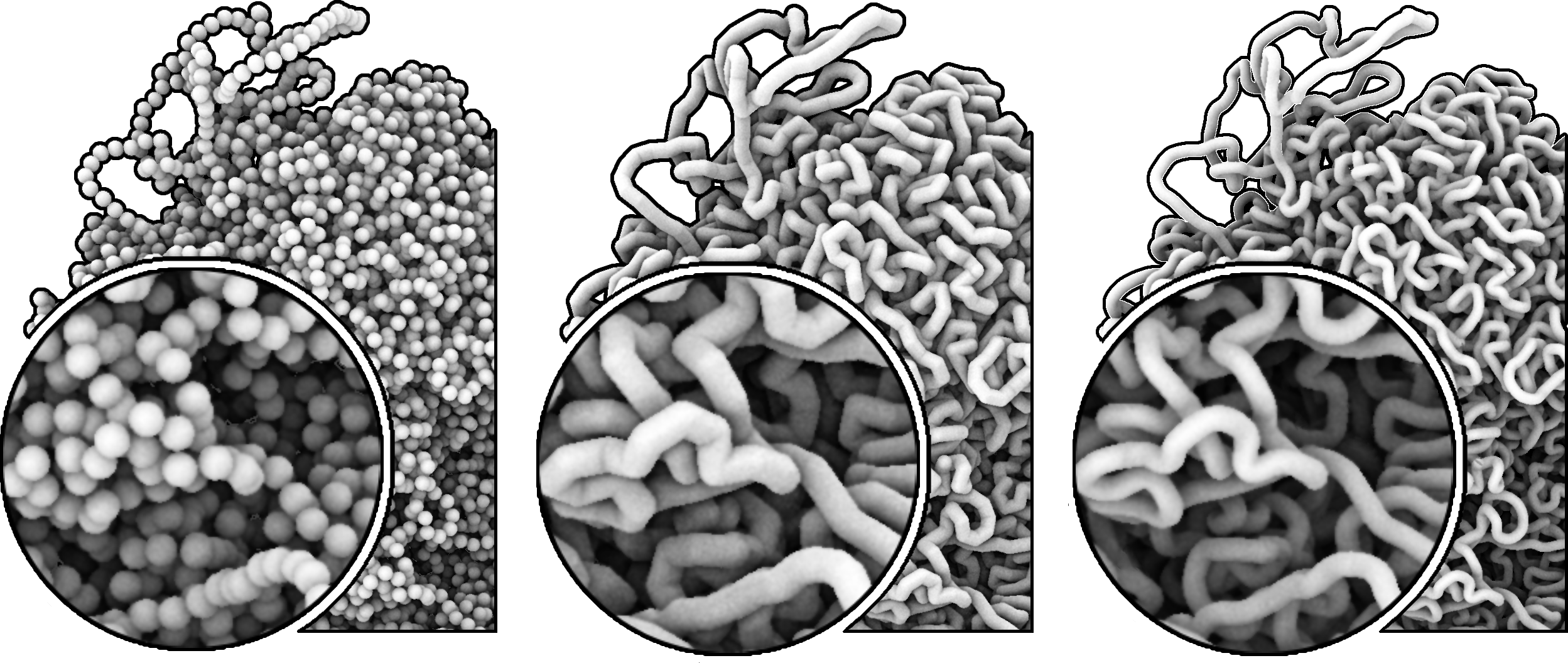}
  \vspace{-3pt}
 \caption{Three types of visual representations available in \chromoskein. From left: spherical, straight tubular, and smooth tubular representation.}
 \label{fig:ObjectRepresentations}
  \vspace{-10pt}
\end{figure}

\changed{3D chromatin modeling methods output a set of discrete 3D positions associated with nucleotide bins.
The set is ordered, \ie, the order of bins corresponds to the order of individual segments in the DNA sequence.}
% Some visualization tools display this data directly as a set of spheres.
% However, chromatin fiber is a continuous strand.
% Consequently, for some tasks, domain experts wish to preserve the continuity information, \ie, to see the order of bins.
%
\changed{We reviewed representations used in the existing 3D chromatin visualization tools, see~\autoref{tab:RW}. In the majority of cases, tools include spherical and tubular representations, sometimes combining both in what is known in molecular visualization as balls-and-stick representation. Some tools implement more unconventional marks, such as crosses in Nucleome Browser~\cite{zhu:2022:nucleome-browser}.}
Working with large 3D models, visual clutter is of primary concern. For that reason, simple and clear visual marks are preferred.
%% \dk{The cross representation leads to a high degree of visual clutter.} %% DK: probably move someplace else
%% \dk{Visual clutter can also be used as a reason not to include balls and stick: you are adding twice as many marks!}
%
\changed{We decided to include} three established spatial representations---a \textit{spherical} and two \textit{continuous tubular} representations, see Figure~\ref{fig:ObjectRepresentations}.

\changed{The \textit{spherical representation} is the most straightforward way of displaying the raw modeling data. Each bin---representing a certain number of nucleotides---is mapped to a sphere. This visual encoding indicates that the positional information is available only at this granularity for this whole genomic segment, \ie, it is impossible to say where each nucleotide lies in 3D space.}
The lost connectivity information is an undeniable limitation of the spherical representation. Therefore, we also provide two continuous representations. In the first, \textit{straight tubular} representation, the consecutive bin positions are connected with straight tubes. This representation communicates the connectivity, while the original data points can still be inferred from the positions of the tube joints. However, the results can look unnatural, particularly for low-resolution data, where long straight segments and sharp angles at joints occur. 
%\tochangenote{image to show this? maybe in supplemental materials!!!}
%
Therefore, the last, \textit{smooth tubular} representation, uses the bin positions to define a spline that forms the centerline of a tube. This technique produces more visually pleasing and organic-looking results, but it comes at the cost of no longer providing precise information about original bin positions.

%% \dk{TODO: discuss the choices and alternatives: why no balls and stick, why this is the optimal selection of design...}
\changed{We chose these three options mainly because of their simplicity. We specifically decided not to include a balls-and-stick representation: 
In this visual mapping there are twice as many marks on the screen, which leads to visual clutter without any added benefit.}

\subsection{3D Rendering \& Shading} %% how we render those visual marks
\changed{We currently deal with chromatin spatial models containing over tens of thousands of bins: our largest model of Mouse genome comprises of 25 thousand bins. }
%
%% When it comes to rendering the spatial structure of chromatin, we currently need to deal with \tochange{over tens of thousands of objects}. %% DK: make more of an analysis: a) genome size, b) common bin resolution
In the future we expect even bigger models, as experimental methods improve in resolution.
\changed{Such data sizes present a challenge for rendering in real-time, especially in the web environment.}
To achieve high performance, we avoid polygonal representation and instead use billboards with screen-space ray-traced parametric objects. The advantage of this approach is twofold: We lower the number of vertices that need to be updated every frame and we obtain a pixel-perfect objects representation. 
For the straight tubular representation, we use efficient and visually correct rounded cones described by Groß and Gumhold~\cite{Lines:2021}.

To compute the smooth tubular representation precisely, we would need to fit a cubic spline through the data bins. However, to achieve fast rendering, we only approximate the cubic spline with quadratic Bezier curves, as proposed by Truong et al.~\cite{CubicBezierInterpolation:Yuksel}. To render the tube itself, we use the method by Reshetov and Luebke~\cite{PhantomIntersector}.

When rendering the tubular representations, we estimate the optimal tube thickness based on the proximity of bins.
\changed{To ensure a reasonable thickness, we limit it to half
the distance between two bins so that two bins do not overlap
and visibly merge into one. Depending on the used model
reconstruction algorithm, the spacing between consecutive bins is not always uniform.
%% and may contain outliers.
Therefore, we use a statistical rule based on the interquartile range (IQR), to bound it between: $Q_1/Q_3 \pm 1.5*IQR$, where $IQR = Q_3 - Q_1$, $Q_1 =$ lower quartile, $Q_3 =$ upper quartile. The default thickness is set in the middle of this range and users can adjust the value within the range.}

\changed{When it comes to shading, virtually all the existing chromatin visualization tools we reviewed employ a simple Phong shading model.
%% ~(\autoref{fig:SSAOComparison}a).
Some tools even skip all shading and only show the 3D model with flat colors~\cite{djekidel:2017:hic-3dviewer}.
Phong shading model considers only local illumination omitting shading by neighboring structures. For larger chromatin models, it leads to an ambiguous view where it is impossible to discriminate global features such as holes and crevices. Thus, the comprehension of an overall shape is hindered.}

\begin{figure}[b!]
    \centering
	\setlength{\subfigcapskip}{-4ex}%
	\textcolor{black}{\subfigure[\hspace{.48\linewidth}]{%
	\includegraphics[width=.48\linewidth]{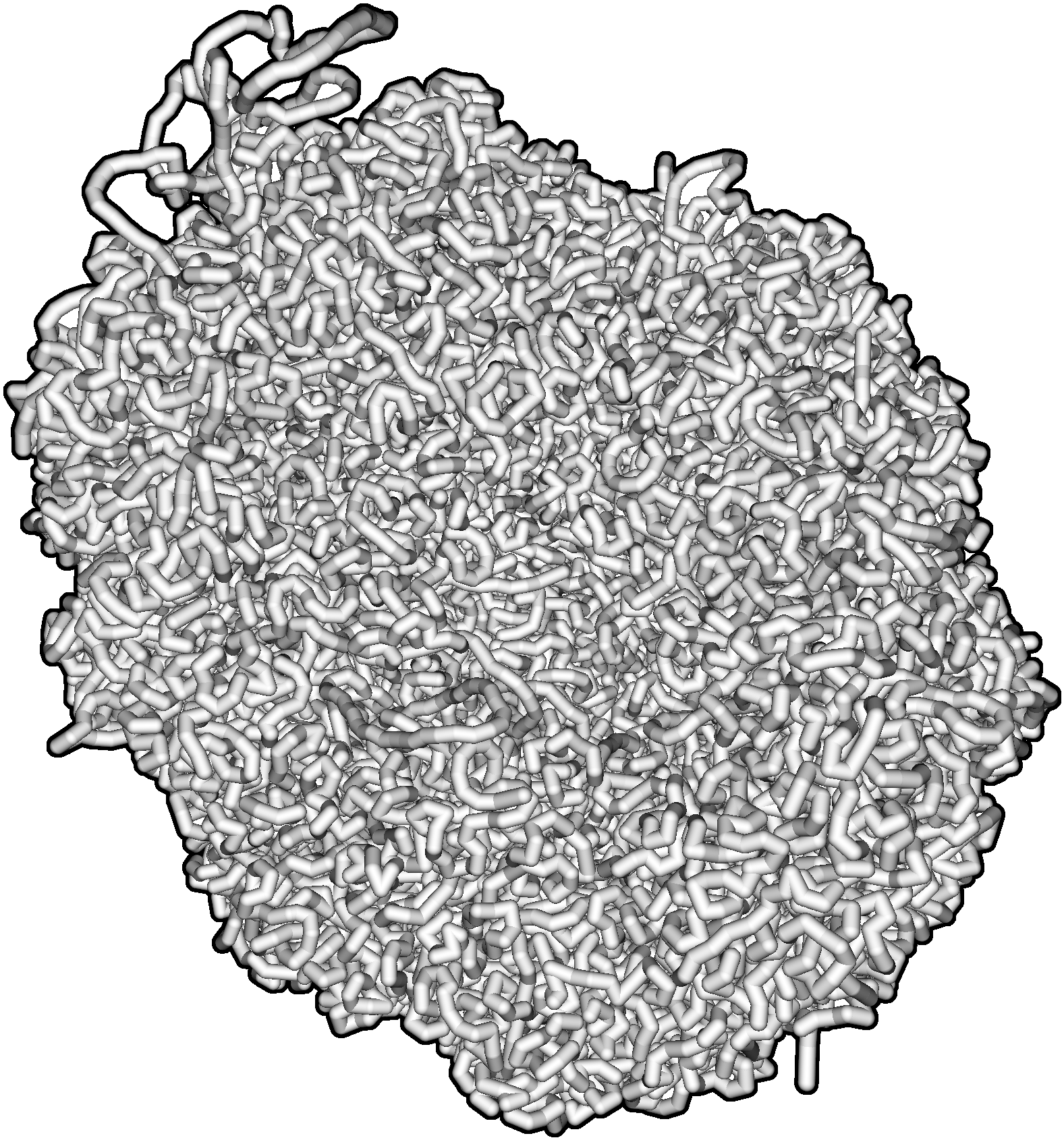}%
	\label{fig:ssao-phong}
 }}\hfill%
    \textcolor{black}{\subfigure[\hspace{.48\linewidth}]{%
	\includegraphics[width=.48\linewidth]{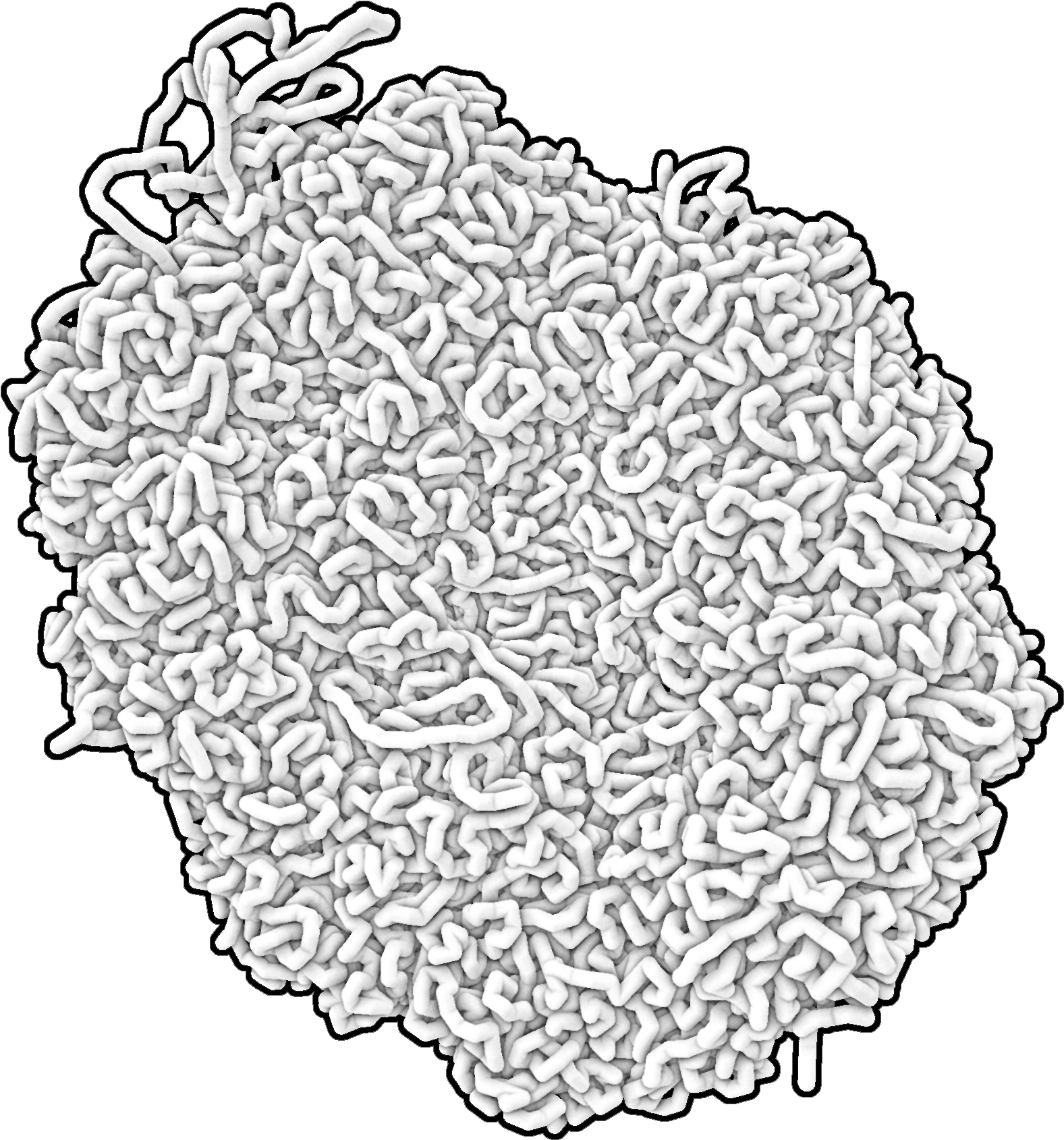}
	\label{fig:ssao-near}
}}
    \textcolor{black}{\subfigure[\hspace{.48\linewidth}]{%
	\includegraphics[width=.48\linewidth]{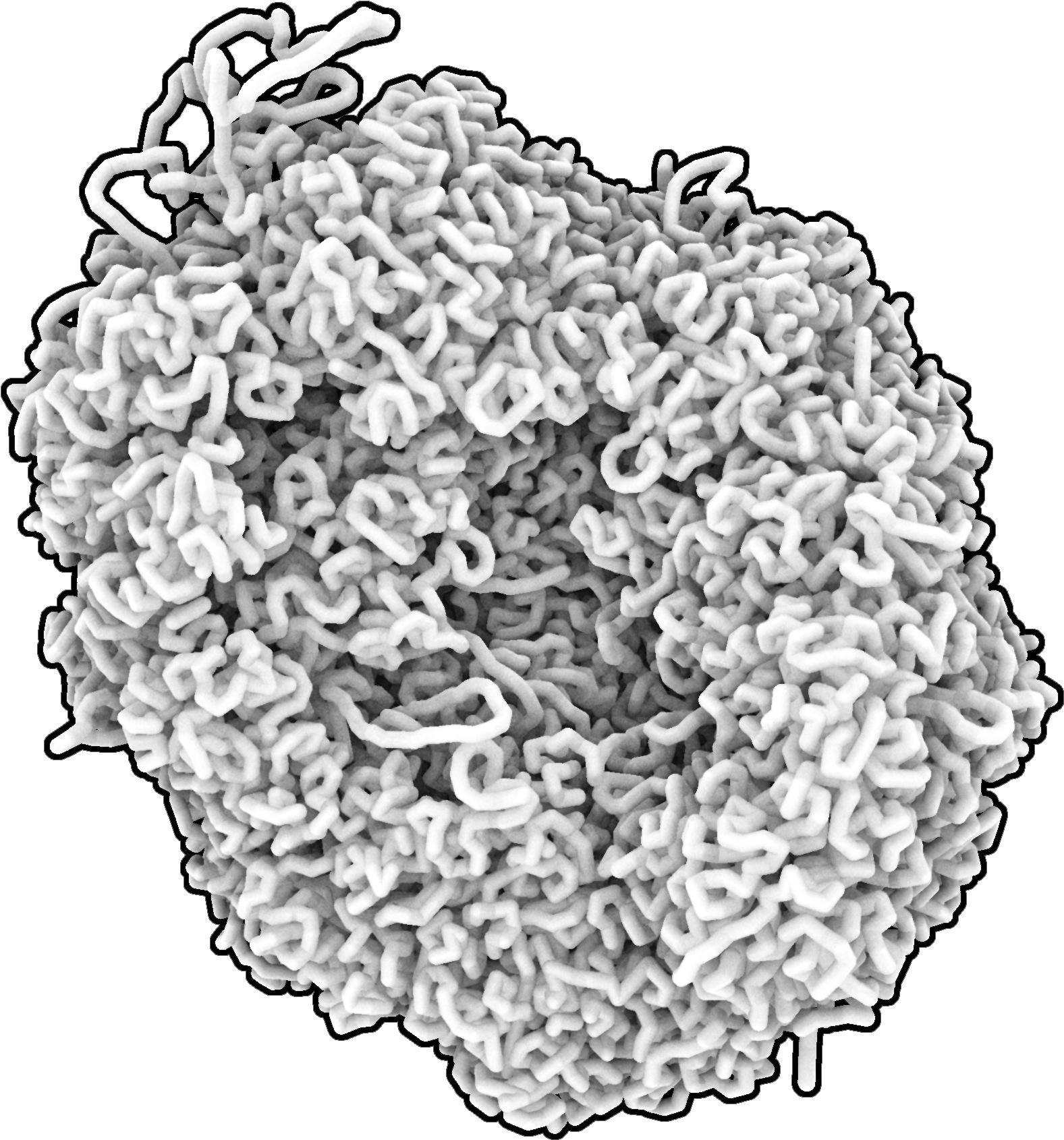}%
	\label{fig:ssao-far}
}}\hfill%
\textcolor{black}{\subfigure[\hspace{.48\linewidth}]{%
	\includegraphics[width=.48\linewidth]{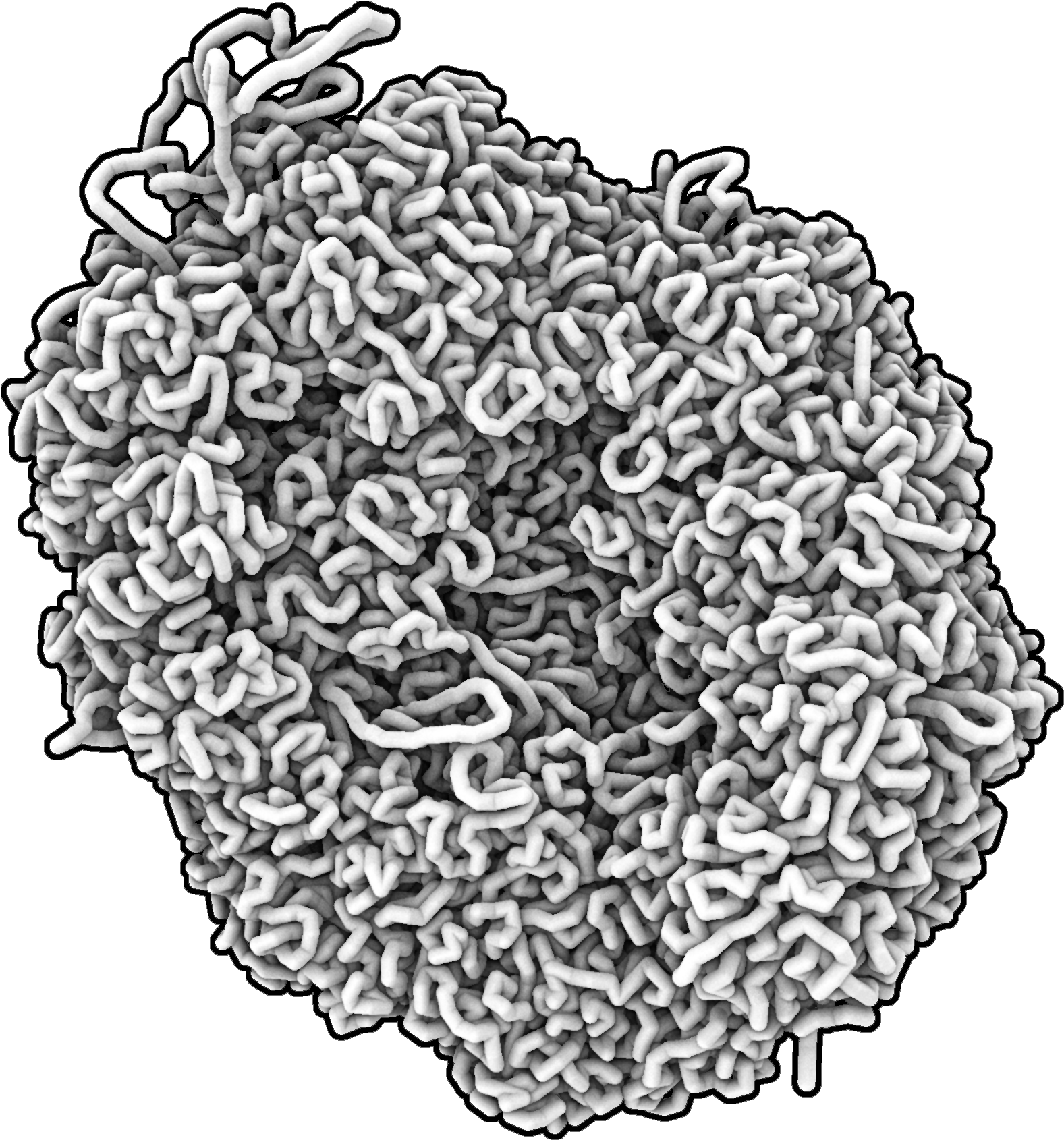}%
	\label{fig:ssao-combined}
 }}
 \vspace{-1ex}

    \vspace{-3pt}
    \caption{Shading greatly influences perception of the overall shape. The standard Phong shading model results in a highly cluttered view~\subref{fig:ssao-phong}; SSAO with a small radius highlights local features~\subref{fig:ssao-near}, while a bigger radius accentuates global features~\subref{fig:ssao-far}. We use two SSAO passes with different radii to combine benefits of both~\subref{fig:ssao-combined}.}\vspace{-1ex}
    \label{fig:SSAOComparison}
     \vspace{-3pt}
\end{figure}

To enhance perception of such spatial features, we shade the scene with real-time screen space ambient occlusion (SSAO)~\cite{tarini:ssao:2006}. It can be difficult to configure the SSAO radius parameter for large datasets, as a large radius accentuates only bins deeply inside the structure. In contrast, a small radius highlights only differences between close objects. We rectify this by stacking two results of SSAO computations with small (for near objects) and large (for deeply buried objects) radii. 
Compared to the Phong shading model,
%% utilized in most of the existing 3D chromatin visualization tools,
in \chromoskein~we are clearly able to comprehend shape features, such as a hole in the center of a genome occupied in cells by nucleolus. A comparison can be seen in Figure~\ref{fig:SSAOComparison}.

\subsection{\changed{Interactions With 3D Chromatin}} %% what we can do with the rendered marks
%% Both are kinda connected (chicken and egg): you might want to select something ''inside'', so you need occlusion management; and one occlusion management is semantic toggle, which needs to know about selections.
%% \changed{The visualization of 3D models in \chromoskein~can by itself be useful to bioinformaticians. However, our primary goal was to allow interaction with the 3D data.
\changed{We aim to give bioinformaticians the tools necessary to go beyond static visualization and to support them in analyzing the 3D model directly in its spatial context.
Next, we describe two interconnected interactions used to filter, explore, and examine the 3D data.}

\subsubsection{Selections}
\label{sec:selections}
\newcommand{\PointSelection}{\includesvg[width=8pt]{figures/icons/point.svg}~\texttt{Point selection}}
\newcommand{\SegmentSelection}{\includesvg[width=8pt]{figures/icons/linear.svg}~\texttt{Continuous sequence selection}}
\newcommand{\SphereSelection}{\includesvg[width=8pt]{figures/icons/lasso.svg}~\texttt{Sphere selection}}

\begin{figure}[tb]
 \centering
 \includegraphics[width=\columnwidth]{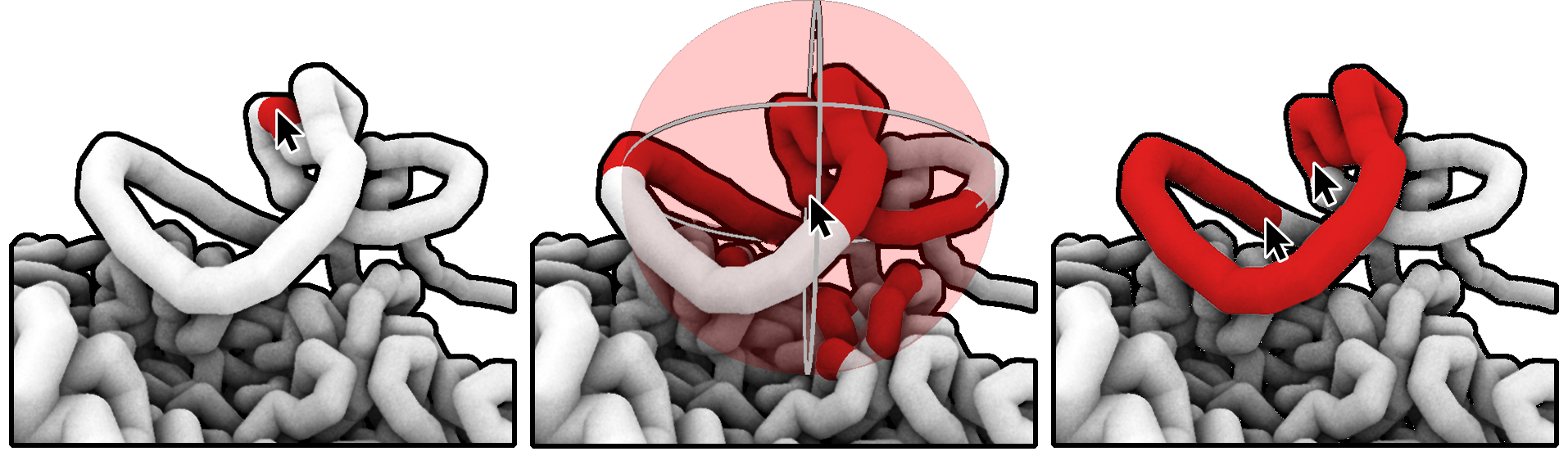}
 \vspace{-6mm}
 \caption{From left to right: \PointSelection, \SphereSelection, \SegmentSelection.}
 \label{fig:selection-types}
\end{figure}

%% \annot{TODO: sprinkle some Nusrat here}
\changed{One of the key interactions with 3D chromatin required by biologists is the ability to select genomic regions.
The selection task is a prominent interaction across all visualization tools~\cite{wills:96:selections}.}
\changed{Available genomic tools, for the most part, implemented selections in 1D and 2D representations and highlighted the selected parts in the 3D model. Very few of them allow selecting objects directly in the 3D model.}
\changed{A bioinformatician might wish to select bins based on some spatial feature, \eg, bins close to a surface or, the opposite, deeply nested in the nucleus. Such task is impossible to do in the matrix representation. Therefore, in \chromoskein, we designed methods for selecting bins in the 3D view.}

%% \annot{selection = set of bins}
\changed{In the 3D view, the atomic element is a single bin. Thus, we choose to represent selections as a subset of the model bins. }
\changed{We allow users to create an arbitrary number of selections.} 
\changed{
%We encountered several types of selection tools in the existing applications.
In \chromoskein, we implemented three selection tools for the 3D view that together support all possible interactions we identified as critical for the users: \PointSelection, \SegmentSelection, and \SphereSelection, all demonstrated in~\autoref{fig:selection-types}.}

The simplest way of selecting bins is selecting one bin at a time. \PointSelection~serves for the definition of small selections of visually prominent bins
%% , \eg, protruding structures or bins with high values of color-mapped tracks,
or fine selection refinements. It is performed by clicking on the given bin in the 3D view.
One of the most important features found in 3D chromatin is
%% the proximity information, \ie,
which bins or genomic loci are located in a neighborhood to a selected location. Therefore, \SphereSelection~allows selecting bins within a certain spatial distance from a given point or bin.
%% To define this type of selection,
Users can again pick a single bin by clicking on the 3D representation, but in this case, all bins within a spherical radius are selected. The radius is adjustable and the bins within the radius are highlighted upon hover.
% , as can be seen in the middle part of~\autoref{fig:selection-types}.
%
Finally, biologists are sometimes interested in chromatin loops, where the loop end points are distant in the linear sequence but spatially close.
To allow selecting loops in 3D, we use \SegmentSelection.
When two bins are picked in the 3D view, the linear sequence of bins between them is selected.
\changed{We display the selections in 3D by coloring the visual marks representing bins. The colors are randomly generated for each track but can be adjusted by the user.}
% \mt{See Figure~\ref{fig:representations-highlight}.}

\subsubsection{Occlusion Management}
\changed{Chromatin fiber is tightly packed in a nucleus, resulting in a highly intertwined structure with large parts hidden due to occlusion.
%% Occlusion hinders exploration of the model.
Biologists are interested to know, for example, how deeply nested a genomic locus is---the position can affect the function or regulation of genes located in the locus.
Therefore, providing bioinformaticians tools to manage occlusion is essential.
We include two techniques to filter the model by adjusting visibility of its parts: one semantic and one based on spatial position.}

%% \textbf{Semantic Visibility Toggle:}
%% \changed{The first fundamental tool for adjusting visibility is \textbf{Segmentation Visibility Toggle}.
\changed{\chromoskein~allows users to segment the chromatin model by adding bins to selections tracks. The first tool for filtering the 3D model is by \textbf{toggling visibility} of each such segmentation track. Users can hide any track which leads to hiding of associated bins from the 3D representation.
Segmentation tracks can also be loaded from an external file.
%% We describe later in~\autoref{sec:color-by-segmentation}, that the model can also be segmented based on an external file.
This allows, for example, loading chromosome segmentation and exploring the model by toggling visibility of individual chromosomes.}

%% \textbf{Cutting Planes:}
\changed{Sometimes the focus of the analysis cannot be defined semantically but rather spatially.}
%% For example, when users are just getting familiar with the data for the first time and want to get a general overview of the structure's shape.
% To support these cases, \chromoskein~provides another commonly used technique---\textit{clipping planes}.
To support these cases, \chromoskein~provides \textbf{cuttings planes}.
Users can add one or more cutting planes either along one of the major axes or an arbitrarily oriented plane spanning from the camera's point of view.
The model is then \emph{cut} by the plane: The primitives building the 3D bin representations are clipped at the intersection and resulting surface holes are filled.
%and new surfaces are created to fill the holes.\tochangenote{Matus: rewrite this better?}}
%

\begin{figure}[tb]
 \centering
 \includegraphics[width=0.85\columnwidth]{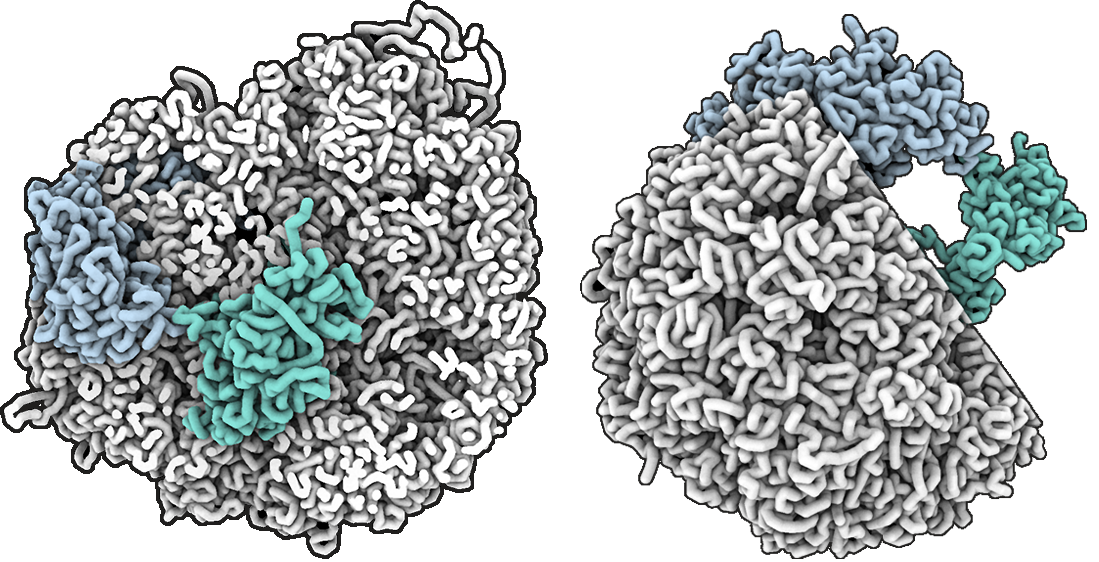}
 \vspace{-7pt}
 \caption{Demonstration of our cuttings planes. Entire chromatin is cut by a user-defined cutting plane and holes are filled. Two users' selections (chromosomes) are configured to be unaffected by cutting planes.}
 \label{fig:cutting-planes}
 \vspace{-5pt}
\end{figure}

\changed{Clipping planes work in tandem with our rendering style supporting depth cues, highlighting both local and global structural features, such as holes nested inside the model.}
\changed{Users can also choose to keep some selections visible at all times which can be useful for studying immersed parts while keeping parts of their surroundings.
All these features are presented in Figure~\ref{fig:cutting-planes}.}

%% \dk{More advanced techniques can be implemented in the future: For example, we envision that exploded view approaches will allow inspecting deeply nested structures without removing any parts of the model.}
%% \dk{Alternatively, magic lens methods might also provide the means for peeking into the structure.}
%% % but go beyond the initial design of our tool.}
%% \dk{However, \chromoskein's two occlusion management tools provide the necessary basis for examining the 3D model.}

%% visualization can be used to keep all elements in scene but still allow diving inside deeply nested structures. For the purposes of this initial tool description, we however consider it above the bare necessary tools.}

%% \section{Linking 3D \& Other Views (KF)}
\section{Linking 3D With Other Data}
%% \changed{3D chromatin models are instrumental to posing hypotheses related to the spatial conformation of genomic features. By itself, however, the 3D model does not give the full picture.}
\changed{Due to the complexity of investigating organization and function of cell nucleus---brought by the dynamicity and scale range---it is essential that biologists integrate several available modalities, each focusing at different aspects of chromatin fiber.}
%
% \dk{Other genomic data sources have been used to investigate chromatin function. Many visualization types and systems have been proposed for working with this data, as reviewed by Nusrat et al.~\cite{nusrat-2019}.}
\changed{Other genomic data and visualizations have been proposed, as reviewed by Nusrat et al.~\cite{nusrat-2019}.}

\changed{In this section, we dive into the design of linking the 3D representation described above with non-spatial data types and their conventional visual representations.}

\subsection{Linked Views}
\begin{figure}[t!]
 \centering 
 \includegraphics[width=0.95\columnwidth]{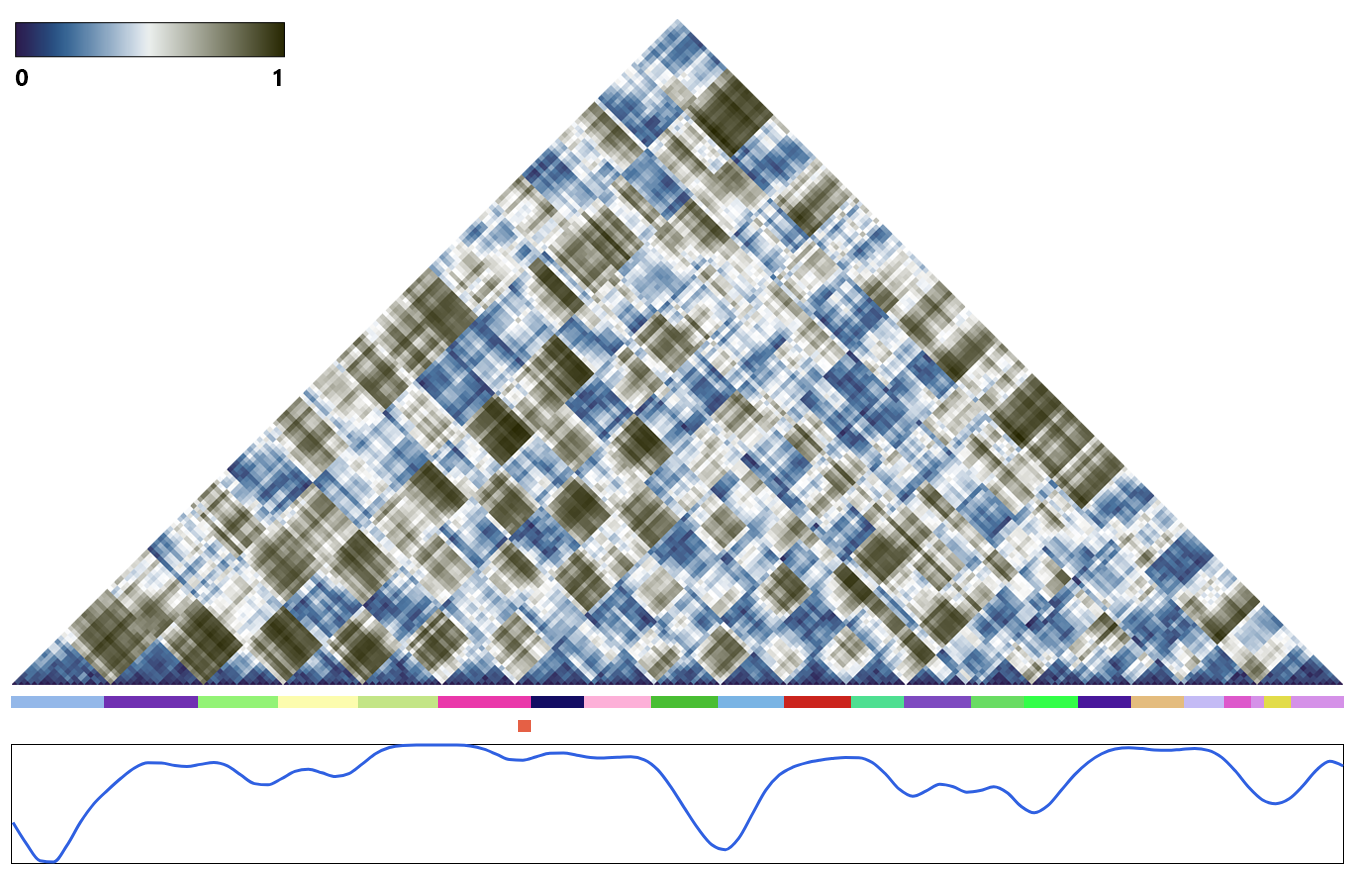}
  \vspace{-5pt}
 \caption{Example of a track view containing 4 tracks (from top to bottom): distance map, segmentation into chromosomes (loaded as selections from a file), custom user-made selection in 3D viewport, and 1D track (in this case solvent accessible surface area).}
 \label{fig:distance-and-genomic-signals}
  \vspace{-10pt}
\end{figure}

\changed{Visualization systems typically connect different data by using multiple \emph{linked views}~\cite{munzner:2014:visualization}.}
% This technique builds on the concepts of juxtaposition, \ie, placing two or more elements side by side.
% In interactive systems, and for data that are somehow related, this design is often accompanied by \emph{linked highlighting}
% %% , also called \emph{brushing and linking},
% that enables users to select data points in one view and reflect this selection in the other views.
% In this design, the 3D and non-spatial views exist separately in colocated view panels and are linked through interaction.}
%
\changed{Not all tasks are best performed in 3D view. Additionally, many biologists are already trained on and used to 1D and 2D visualizations. Therefore, }
% In order to demonstrate linking the 3D representation with other data types and visualizations, 
we include two conventional genomic visualizations in \chromoskein.
First, we implemented a \emph{genomic browser} to display 1D genomic signals. Second, we included a \emph{distance map} to serve as a proxy for any 2D maps, \eg, Hi-C matrix viewer.
\changed{In \chromoskein, we couple these two linear data visualizations into a \emph{track view}, shown in \autoref{fig:distance-and-genomic-signals}.} 

\changed{Both the distance map and the genomic browser are prototypical implementations to complete the feature set required by the bioinformaticians analytical workflow.}
\changed{There are tools that focus more on each modality, \eg, HiGlass for Hi-C maps, but do not include \chromoskein's capabilities for 3D data.}
% \dk{Furthermore, they represent an auxiliary type of non-spatial visualization panels. In the future, we envision that other custom-designed visualizations will be added to \chromoskein. In this initial paper, we purely focused on defining the basic functionality.}

\subsubsection{\changed{Track View: Genomic Browser}}
\changed{\textbf{Genomic browsers} are the most common way of working with genomic data. They display data laid out along one axis---usually the horizontal \emph{x-axis}---annotated by genomic coordinates.}
\changed{Two typical examples of data shown in genomic browsers are \emph{genomic signals} coming from biological experiments, \eg, ChIP-seq which analyzes protein-DNA interactions~\cite{johnson:2007}, and \emph{gene annotations} marking positions of genes on the DNA sequence.}
%% The output of some methods, such as ChIP-seq (analyzing protein-DNA interactions)~\cite{johnson:2007} is in the form of numerical data for each feature, with dense coverage, while gene annotations are sparse and categorical in nature. Track features can also be used for marking Compartmentalization or marking domains of the chromatin strand.

%% \annot{loading 1d data}
\changed{In \chromoskein, we differentiate two types of 1D data: segmentations and signals. Segmentations assign genomic regions to a segmentation track while signals assign numerical values to each bin.}
%
%% Signals on the other hand assign numerical values to each bin.}
\changed{The signals can be loaded into \chromoskein~in BED format.}
\changed{Segmentation tracks can either come from an external file or result from user selections.}
%% as described in~\autoref{sec:selections}.}
%
\changed{We display signal tracks as simple line charts while segmentations are drawn as horizontally stacked bar charts.}

%% \annot{moved from bg}
%% \dk{The sequence of nucleotides can be annotated with additional information gained from various experimental methods in the form of so-called genomic tracks.  
%% Tracks are split into segments, where each segment, also called a \emph{feature}, is defined by its start and end position in the sequence of nucleotides. 
%% The output of some methods, such as ChIP-seq (analyzing protein-DNA interactions)~\cite{johnson:2007} is in the form of numerical data for each feature, with dense coverage, while gene annotations are sparse and categorical in nature. Track features can also be used for marking Compartmentalization or marking domains of the chromatin strand.
%% Two common formats of genomic tracks are the GFF~\cite{the-sequence-ontology_2020} and BED~\cite{kent:2002:ucsc} file formats. While GFF format is commonly used for gene annotations, BED format is used for arbitrary annotations.}

\subsubsection{\changed{Track View: Distance Map}}
\changed{Hi-C experiments typically output 2D matrices with numerical values denoting frequencies of contact between bins.}
\changed{In \chromoskein, a \textbf{distance map} stands in for a full Hi-C map. The distance map is generated from the 3D model and in principle should contain the same information. 3D reconstruction algorithms are rated according to the difference between the generated distance map and the original Hi-C map. If the reconstruction is good, the two 2D maps should be more or less equal.}

\changed{We display the distance map as a triangular shape. The horizontal axis shows bins and each triangular field carries the distance information mapped to a color value.}
\changed{Users can zoom and pan around the 2D map.}
\changed{As an optimization, the distance map is calculated on-demand based on which part of the map is zoomed into.}

\changed{We employ a simple level-of-detail scheme.}
\changed{If the size of a distance map is such that it cannot allocate enough pixels for all bins, a level change is applied.
%% level-of-detail technique is applied.
We merge multiple consecutive bins by averaging their positions and calculate the distance between those larger bins. }
\changed{Distance maps can be also generated from custom selections. This can be useful to observe distance relationships of a subset of chromatin.}

% \dk{The 3D view, the 1D genomic browser, and the 2D distance map are three types of views that users can currently use in \chromoskein. Next, we discuss the way these views are linked.}

%% \annot{moved from bg}
%% As mentioned above, Hi-C experiments output square contact matrices of numerical values. 
%% Additionally, similar information can also be provided in the form of 2D \emph{distance maps}, which substitute contact frequencies with spatial distances between bins of predicted 3D model of chromatin. Since the 3D model is based on the original Hi-C data, the contact frequency in Hi-C maps is highly correlated with the distance,
%% and in many cases, these two measures can be used interchangeably.
%% The 2D matrices can be given in standard generic formats, such as CSV, or in formats designed for Hi-C maps, such as \texttt{cooler}~\cite{cooler:2021}.

\subsubsection{\changed{Interactive Linking}}

\changed{Nusrat et al.~\cite{nusrat-2019} mention that views can be weakly, medium, or strongly linked. That simple categorization applies for 1D or 2D views that share reference frame of genomic coordinates. The 3D view steps out of this conformity, which extends the design space of possible linking.}

%% \annot{many panels}
\changed{We allow users to create many panels of the two types, \ie, 3D view or track view.}
%% \changed{They can also decide if some views are linked or not.
\changed{Users can synchronize cameras between 3D views.
That way they can observe one 3D model with different adjustments, \eg, hidden specific chromosomes.}
%% That way they can observe one 3D model annotated by different attributes at the same time.}
%% That way they can easily see multiple signals mapped on the structure and select areas of interest derived from correlation of those 3D views.}
%
\changed{The 2D distance map and 1D genomic tracks in a track view are linked by default: Zooming and panning will be reflected across all tracks in one track view.}

%% DK: from 2D
\changed{The linking between a 3D view and a track view is accomplished through custom selections.}
%% \changed{As we mentioned above,
\changed{In the 3D view biologists
%% need selection tools to
select parts that are interesting because of their spatial features. On the other hand, they might also want to let the selections be guided by either the linear data, \eg, signal peaks, or patterns in the 2D map. For that reason, we implemented several selection tools also for the track view.}
%% 2D map representation.}

%% \changed{This functionality fulfils the requirements of bidirectional interaction. Users can both select salient features in the 3D using tools described in~\autoref{sec:selections}, and also make selections based on features (\eg, signal peaks) in the track view. Both actions will be reflected in the other representation.}

\subsection{\changed{Annotating the 3D Structure}}
\begin{figure}[tb]
 \centering
 \includegraphics[width=\columnwidth]{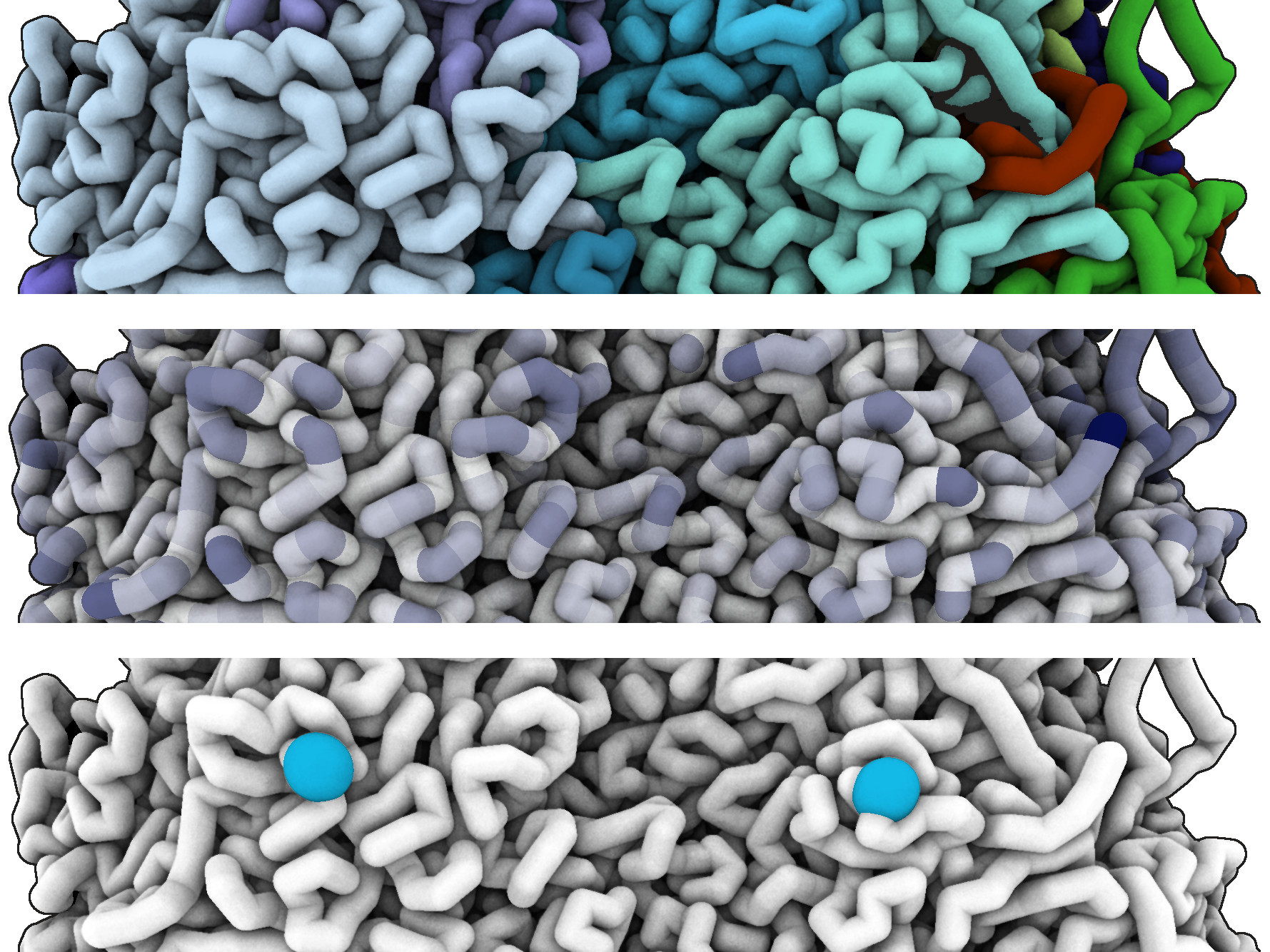}
 \vspace{-5mm}
 \caption{From top to bottom: chromatin colored by segmentation (chromosomes), genomic signal (solvent accessible surface area), markers.}
 \label{fig:annotations}
\end{figure}

\changed{While linked views are based on the concepts of \emph{juxtaposition}, \emph{superimposing} data in the 3D is also possible.}
%% can also be used. Here the data are overlaid over the 3D view.}
\changed{This allows biologists to contextualize 1D genomic data in space. }% or highlight partitioning of the chromatin fiber.}
%% \changed{Already many existing 3D chromatin visualization tools use superimposition to encode additional data. The most common case is to denote a segmentation of a model into chromosomes by different colors.}
\changed{Biologists call this \emph{annotating} the 3D model.
There are several ways that the color channel can be used to augment the 3D structure with other information.}
\changed{All types of annotation are demonstrated in \autoref{fig:annotations}.}

%% \subsubsection{Coloring By Segmentation}
%% \label{sec:color-by-segmentation}
\changed{The first type of annotation is coloring the 3D representation based on chromosome segmentation, as seen in several existing 3D chromatin visualization tools.
However, other segmentations can also be relevant: \eg, A/B compartments or TADs.
In \chromoskein, we therefore generalize this type of annotation.
We allow users to segment the model by selecting its parts using several selection tools
% \dk{Alternatively, this segmentation can come from an external file.}\tochangenote{BED format?}
% \dk{In such a case, each bin carries a binary value of belonging to the segmentation track or not.}
or loading the segmentation from an external file.}
\changed{We then color the 3D representation as we described in~\autoref{sec:selections}.}

%% \subsubsection{Coloring By Genomic Signals}
\changed{However, annotating the 3D model based on segmentation is not the only way to use the color channel. We can also map genomic signals from other biological experiments and superimpose the 1D data onto 3D.}

\changed{These types of data typically have more fine-grained resolution than 3D structures. Therefore, we allow users to choose how multiple values per bin are mapped, \ie, a minimum, maximum, average, or median.
In order to color bins, we normalize the data and then apply one of the color maps recommended by the scientific color guide~\cite{scientific_color_guide}.}

%% \annot{coloring by computed properties}
%% DK: this is actually quite important if we're using SASA in use cases, here we should mention it
% \dk{
\changed{In some cases, the values might not come from an external file but instead can be computed on-the-fly.
Bioinformatics tools often require sophisticated computation and powerful hardware but some algorithms can be implemented and run directly on a client device.
This removes the need for running an external tool and importing the results into a visualization system. Users can thus iterate faster and often re-define the model subset meant for computation.
As an example, we include computation of Solvent Accessible Surface Area (SASA) in \chromoskein.}

% An example is the computation of Solvent Accessible Surface Area (SASA). While this value can be computed using external tools and imported into \chromoskein, the ability to compute and visualize this value ad-hoc can be valuable when the user wishes to iterate and often re-define the subset of the model on which this computation should be performed.
% }

% \dk{In some cases, the values might not come from an external file but instead can be computed on-the-fly. An example is the computation of Solvent Accessible Surface Area (SASA). While this value can be computed using external tools and imported into \chromoskein, the ability to compute and visualize this value ad-hoc can be valuable when the user wishes to iterate and often re-define the subset of the model on which this computation should be performed.}
% \ki{Therefore, a variation of Shrake \& Rupley algorithm \cite{shrake_rupley:1973:sasa} is implemented in our application.}

%% \subsubsection{Using Markers}
\changed{Finally, biologists are often interested in highlighting short segments of the DNA that carry functional elements such as genes.}
\changed{These short segments, called \emph{loci} (singular locus), generally fall into a range less than a single bin. Therefore, semantically, we can consider them as short segmentations and implement them by selections. However, due to convention, we chose to highlight such loci using \emph{markers}. Since markers most often span just a single bin, we display them as spheres and make them salient by using a greater radius and a different color, adjustable by the user.}
% While semantically these could be categorized as short segmentations and implemented by selections, we chose a special visual representation for this type of data. We call these \emph{markers}, and display them as spheres in the 3D view.}

\section{Implementation}
\chromoskein~is implemented as a client-only web application, which makes it usable without the need to install desktop software. The omission of a remote server component ensures that potentially confidential data need not be uploaded to a third-party server.

\chromoskein~itself is divided into two main parts: the application itself, written using the popular \emph{React} front-end framework, and a narrowly focused visualization library written using the \emph{WebGPU} API~\cite{Malyshau:2022:WebGPU}. We decided to use WebGPU as it offers low-level access to GPU and addresses many of the issues of WebGL. Note that WebGPU is currently in development and requires experimental versions of modern browsers, \eg, \emph{Google Chrome Canary}.
%
% During the development, we focused particularly on the modular architecture, so the application is easy to extend with other viewport types, and with more features in the existing viewport types. For visualization researchers it can therefore serve as a base for prototyping novel genomic visualisation techniques.
%
The project is open-sourced at \texttt{\url{https://github.com/chromoskein/chromoskein}}, along with instructions on how to set up the browser to run our tool.

\section{Exemplary Use Cases}
\begin{figure*}[t]
 \centering % avoid the use of \begin{center}...\end{center} and use \centering instead (more compact)
 \includegraphics[width=\textwidth]{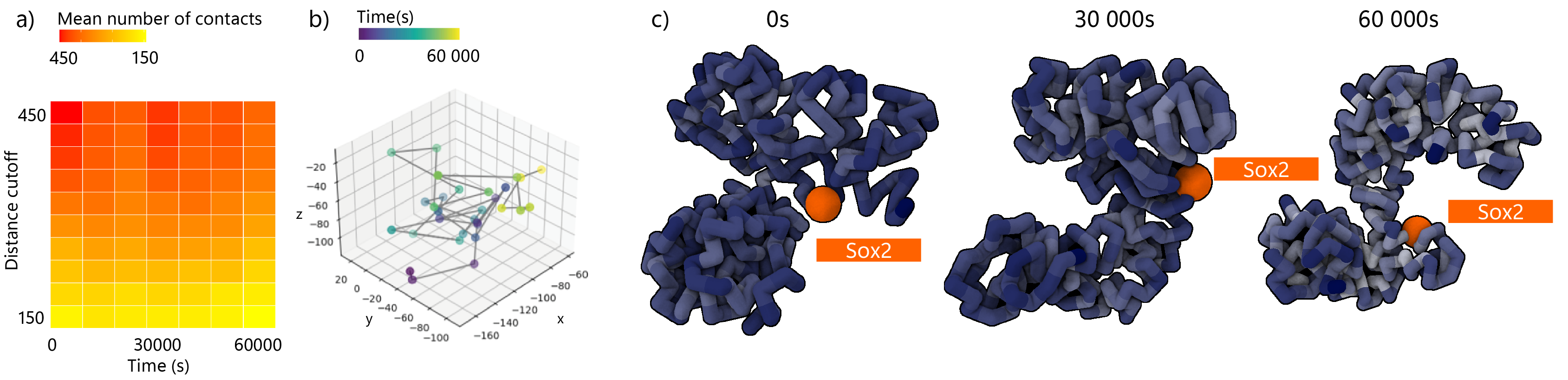}
 % \vspace{-10pt}
 \caption{a) A number of pairwise contacts to Sox2 as a function of time and distance cutoff. b) Measurement of Sox2 trajectory in 3D space. c) Selected conformations of Sox2 locus at time points 0-30,000-60,000s with Sox2 colored in orange. The chromosomal locus is colored based on solvent accessible surface area (SASA) with dark blue corresponding to a higher value while light blue corresponds to a lower value of SASA. Parts a) and b) were made in R programming language. }
 \label{fig:usecase1}
 % \vspace{-10pt}
\end{figure*}

% \mt{We present two use cases that were made in collaboration with a biologist. The biologist provided figures they are capable of creating, pointing out their deficiencies. Afterwards, they built 3D versions in our tool to see additional information not found in 1D and 2D versions. }

\changed{We present two use cases prepared in collaboration with a biologist to demonstrate the capabilities of \chromoskein. The biologist first performed analysis purely from 2D matrices, pointing out its deficiencies. Afterward, they utilized 3D visualization to gain additional insight beyond what is possible from only 1D and 2D genomic data.}

\subsection{\changed{Case study I}}
%% \changed{In the first case study, we took an example of
% \changed{First
\changed{In the first case study, we look at chromosomal dynamics from Di Stefano et al.\cite{DiStefano2020}, focusing on the 3D model of a region surrounding mouse gene Sox2 and the gene itself.}
\changed{Without the utilization of 3D visualization, one has very limited options for exploring the dynamical behavior of Sox2 gene from pure XYZ coordinates; some of the potential analysis approaches are shown in Figure~\ref{fig:usecase1}a-b.}

\changed{Firstly, the number of pairwise contacts to Sox2 gene was explored as a function of time and number of interactions up to certain distance. In most of the selected distance cutoffs, the number of contacts increases in the first 10 000 steps of dynamical simulation while steadily lowering below starting point until the end of the simulation (step 60 000). This suggests a tight packing of the chromatin fiber around Sox2 gene at the beginning of the simulation where the number of the
  %% pairwise
  interactions is higher. As the simulation progresses, the
  %% chromatin
  fiber around Sox2 gene will likely unfold.}

\changed{Secondly, one can monitor the spatial trajectory of Sox2 gene to
  %% better
  understand its dynamics and mobility. If the monitored position fluctuates in all three axes uniformly, its mobility is considered isotropic.
  %% , while
  Constraints in one of the
  axes
  %% directions (axis)
  hint at the presence of anisotropy. This can be related to the stability of specific conformation, which may play an important role in the regulation of gene transcription. }

\changed{However, without further visualization and exploration of 3D dynamics, it is not possible to infer further details. Visualization of Sox2 (Figure~\ref{fig:usecase1}c) trajectory helps to understand these processes better and clarify the hypothesis based on previous analysis (Figure~\ref{fig:usecase1}a-b). Indeed, it confirms the unfolding of chromatin locus in later stages of simulation (\textgreater $10\,000$ s) and the rearrangement of chromatin folding around Sox2. Moreover, Chromoskein can calculate solvent accessible surface area (SASA), which allows further interpretation of un/folding events. Although Sox2 region converges to conformation with fewer interactions (Figure~\ref{fig:usecase1}a), SASA values of most bins are lower towards the end of the simulation (brighter blue in Figure~\ref{fig:usecase1}c) than in the initial conformation, making it less accessible to potential transcription factors (specific proteins regulating gene activity). This suggests the tendency for more permanent gene expression of Sox2 gene with less flexible regulation.}

\begin{figure*}[t]
 \centering % avoid the use of \begin{center}...\end{center} and use \centering instead (more compact)
 \includegraphics[width=\textwidth]{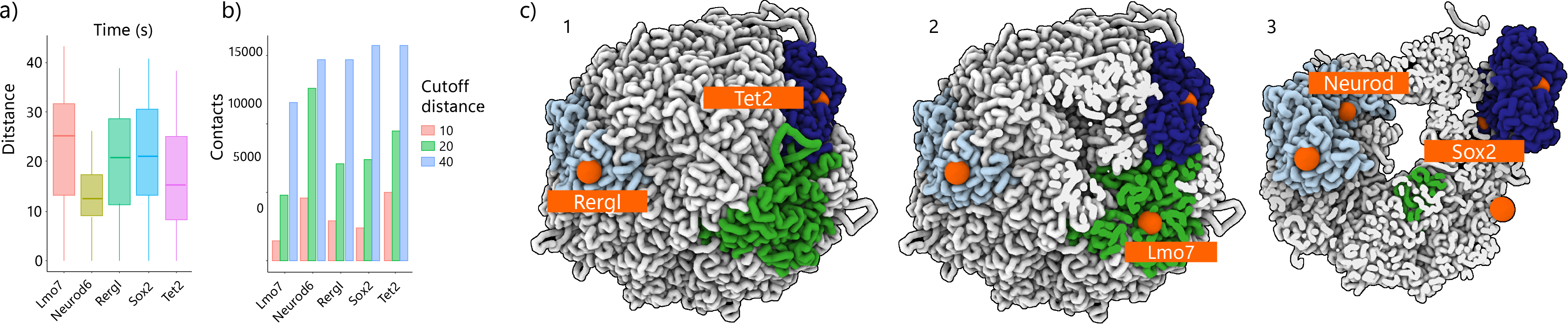}
  % \vspace{-3pt}
 \caption{a) Distribution of pairwise distances from genes of interest (Lmo7, Neurod6, Rergl, Sox2, and Tet2) to the rest of the chromosome where they are located.  b) Number of pairwise contacts for Lmo7/Neurod6/Rergl/Sox2 and Tet2 to the rest of the chromosome where they are located, considering different distance cutoff values (10, 20, 40). c) 3D model of the whole mouse genome with colored chromosomes where Lmo7/Neurod6/Rergl/Sox2 and Tet2 (colored as orange beads) are present. Numbers 1-3 are showing a specific cross-section through the mouse genome in the z-axis. Number 1 only depicts genes localized on the genome surface, while numbers 2-3 reveal additional genes buried inside the chromatin. Parts a) and b) were made in R programming language. }
 \label{fig:usecase2}
 % \vspace{-10pt}
\end{figure*}

\subsection{\changed{Case study II}}

\changed{In the second case study, we inspected the same Sox2 gene in the context of whole genome organization, as published by Stevens et al.~\cite{Stevens2017}, along with the addition of Lmo7, Neurod6, Rergl, Tet2 genes.}

\changed{Exploration of Hi-C matrices or single coordinates without 3D visualization will inform us about probabilistic interactions among all the chromosomes but
not about their mutual spatial arrangements/colocalization. Distance distribution estimates of a selected gene towards the rest of the chromosome can indirectly inform us about the folding of a region surrounding the gene (Figure~\ref{fig:usecase2}a-b). }

\changed{3D visualization of whole genome structure allowed us to uncover that Tet2 and Rergl genes are located on the 'surface' of the genome. Their corresponding chromosomes are likely interacting with so-called nuclear lamina (Figure~\ref{fig:usecase2}c1-c2).
This may suggest genes' participation in specific processes, such as lamina regulation or nuclear transport.}

% Nuclear speckles, nucleolus, and other nuclear non-genome arrangements cannot be captured via Hi-C experiment, thus manifesting themselves within the 3D genome model as cavities or holes inside the genome model.
\changed{There are many non-genome arrangements that cannot be captured via Hi-C experiment. However, they can manifest themselves within the 3D genome model as cavities or holes; therefore, depth cues and usage of cutting planes can provide us with information about them.}

\changed{By sliding a cutting plane through the whole genome 3D structure (Figure~\ref{fig:usecase2}c3), we can uncover the rest of genes, located on chromosomes buried deeply inside the chromatin core. We used the ability to keep chromosomes visible even under the cutting plane to keep the context.}
\changed{We find that Neurod gene is positioned in close proximity to a spatial caveat inside the genome, which is most likely occupied by the nucleolus, suggesting an important regulatory role or potential interaction with rRNA or rDNA.}

% \subsection{Summary}

% \changed{Altogether, this demonstrates the power of 3D genome visualization (both for static and dynamic conformations) in the context of biological interpretation where visualization might often be the crucial missing piece of the puzzle. Uncovering potential biological context via 3D genome visualization allows for the development of new biological hypotheses which can be further experimentally validated.}

% \changed{We also showcased complete combined usage of all features in \chromoskein, depth cues using double SSAO, markers, and configurable cutting planes, to fully utilize 3D genome visualization.}

% \section{TBD: Evaluation / Discussion / Lessons Learned}
% \section{\mt{Lessons Learned}}

% \mt{While implementing \chromoskein and discussing problems, in study of chromatin, with biologists a recurring problem that arose was lack of any standardization in data formats. In 3D, PDB has several limitations when it comes to precision, size limits, compression, and semantic description of chromatin data. Most biologists and tools decided to create their own formats based on CSV. Due to this, none of the 3D tools can be used in parallel. In \chromoskein, we dealt with this by providing CSV import where biologists can tell what kind of CSV the file is and which columns correspond to 3D data. We consider all of this insufficient and an important future problem to be solved before better tools can continue to be developed.}

\section{Conclusion}
We are currently receiving progressively better predictions of how DNA interacts in the 3D space. Formulating hypotheses about overall chromatin conformation intrinsically requires appropriate visual support, which is lacking in the currently available tools. In our work, we aim to bring the 3D spatial representation of chromatin to the forefront and design a novel tool that would assist experts in several stages of their research.
We clearly identified a gap in the available software, where the semantics of the domain problem elicit new methods, making existing molecular visualization tools insufficient. 

We believe that our newly proposed ChromoSkein tool will contribute to understanding of the importance of chromatin spatial organization. We foresee that the description of our design choices, the architecture of the tool, along with the availability of the source code, will serve the communities of chromatin experts, software developers, and visualization experts in adopting our solutions.
%% to their cases.

The described visualization framework gives us a solid basis for further research. Several topics already came up in preliminary discussions with domain experts.
%% dynamics
First, like other biological phenomena, chromatin is a dynamic structure that is constantly changing. Studying these movements is highly important since chromatin adjusts to the cell cycle and drastically changes its conformation.
From the visualization perspective, time series data for 3D models of chromatin present intriguing challenges for which novel solutions will be required.
%comparisons
Second, studying the evolutionary changes in chromatin structure and spatial organization among and across species can be naturally supported by comparative visualization. 
%integration
Finally, chromatin research is intrinsically based on integration of experimental methods, data types, and data sources. The same approach is being applied to tooling, now also enabled thanks to modern web technologies.
In the future, we foresee a high potential in integrating our tool with applications such as HiGlass, that are targeting to solve one specific problem.

% or


\ifCLASSOPTIONcompsoc
  \section*{Acknowledgments}
\else
  \section*{Acknowledgment}
\fi

%% The authors would like to thank...
The presented work has been supported by the Czech Ministry of Education project no. LTC20033.
Authors wish to thank Marc Marti-Renom for consultations and Hanka Pokojn\'{a} for creating \chromoskein's logo.

\ifCLASSOPTIONcaptionsoff
  \newpage
\fi



% \bibliographystyle{IEEEtran}
% \bibliographystyle{IEEEtranS}
\bibliographystyle{IEEEtranS-test}
\bibliography{template}


% \end{thebibliography}


\begin{IEEEbiography}
[{\includegraphics[width=1.0in]{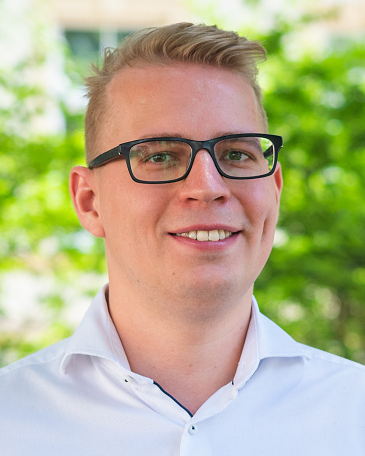}}]
{Mat\'{u}\v{s} Tal\v{c}\'{i}k}
is a doctoral student at Masaryk University in Brno, Czech Republic. He studied computer graphics after which he moved on to visualization. He applies his knowledge to make rendering of large biological data sets beautiful and real-time.
\end{IEEEbiography}
\vspace{-10mm}

\begin{IEEEbiography}
[{\includegraphics[width=1in,height=1.25in,clip,keepaspectratio]{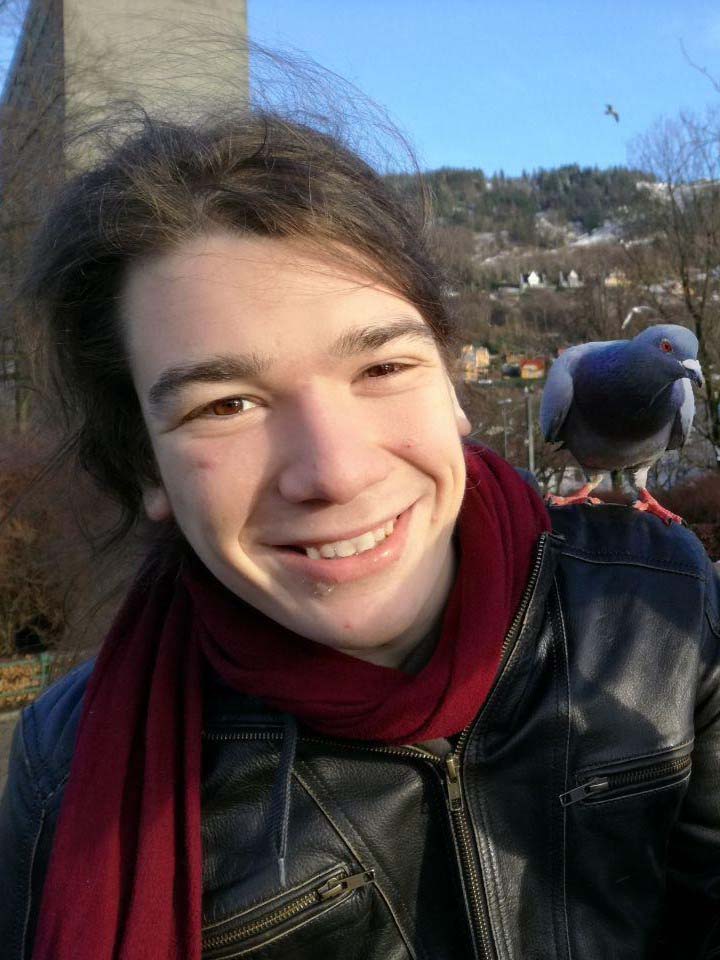}}]{Filip Op\'{a}len\'{y}} is a doctoral student at Masaryk
University in Brno, Czech Republic and a member of Visitlab research laboratory, focusing on vizualization of biological data. 
\end{IEEEbiography}
\vspace{-10mm}

%\newpage

\begin{IEEEbiography}[{\includegraphics[width=1in,height=1.25in,clip,keepaspectratio]{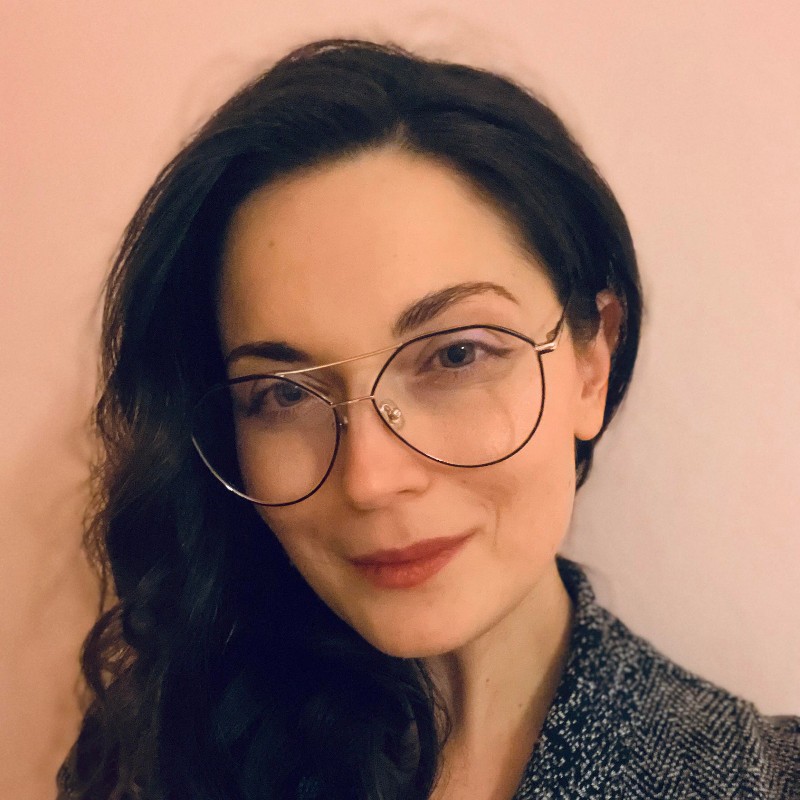}}]{Tereza Clarence}
Postdoctoral researcher at the Center for Disease Neurogenomics at Icanh School of Medicine at Mt Sinai, NY. She received her doctoral degree at the Francis Crick Institute and King's College London, UK in Computational Biology in 2022. Her research focuses on 3D genome function, organization and modelling.
\end{IEEEbiography}
\vspace{-10mm}

\begin{IEEEbiography}[{\includegraphics[width=1in,height=1.25in,clip,keepaspectratio]{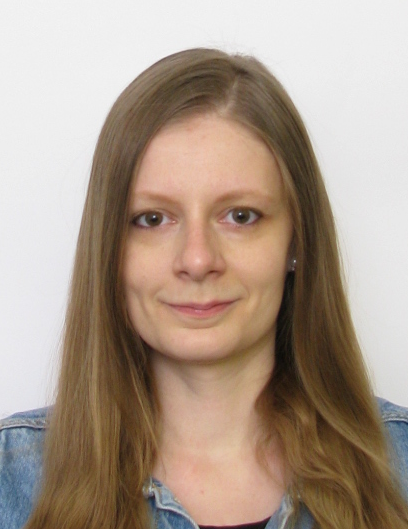}}]{Katar\'{i}na Furmanov\'{a}} is an Assistant professor and member of the Visitlab research laboratory at Masaryk University in Brno, Czech Republic. She obtained her Ph.D. in Computer Graphics in 2019 from the same university.  After finishing her Ph.D., she spent one year as a postdoc at Aarhus University in Denmark. Her current research interest involve visualization of medical and biological data.
\end{IEEEbiography}
\vspace{-10mm}

 \begin{IEEEbiography}[{\includegraphics[width=1in,height=1.25in,clip,keepaspectratio]{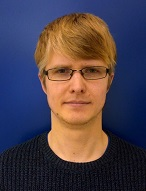}}]{Jan By\v{s}ka} is an Assistant Professor at the Masaryk University in Brno, Czech Republic and a part-time Associate Professor at the University of Bergen, Norway. He is a member of the Visitlab research laboratory, where his work focuses mostly on various challenges in the field of visualization of molecular and time-dependent data.
\end{IEEEbiography}
\vspace{-10mm}

\begin{IEEEbiography}[\href{https://www.fi.muni.cz/~xkozlik/}
{\includegraphics[width=1in,height=1.25in,clip,keepaspectratio]{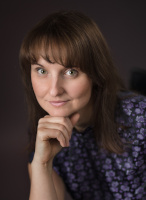}}]{Barbora Kozl\'{i}kov\'{a}}
is an Associate Professor at the Faculty of Informatics at Masaryk University in Brno, Czech Republic. She is the head of the Visitlab research laboratory, specializing in the design of visualization and visual analysis methods and systems for diverse application fields, including biochemistry, medicine, and geography. She has published over 70 research papers.   
\end{IEEEbiography}
\vspace{-10mm}

\begin{IEEEbiography}[\href{http://www.davidkouril.com/}{\includegraphics[width=1in,height=1.25in,clip,keepaspectratio]{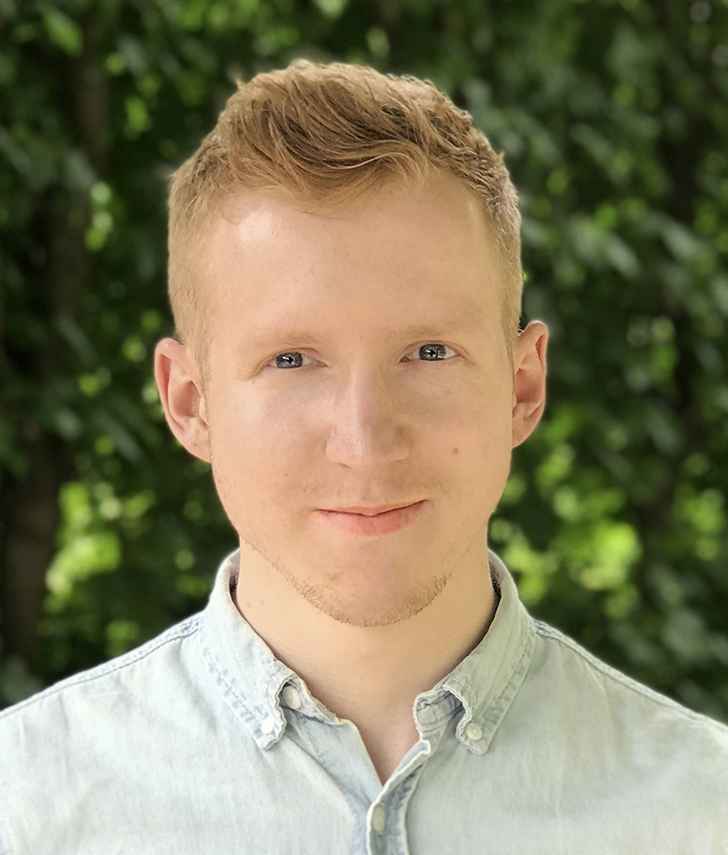}}]{David Kou\v{r}il}
is a postdoctoral researcher at Masaryk University in Brno, Czech Republic. He received his doctoral degree from TU Wien in Vienna, Austria in April 2021. He focuses on three-dimensional biological data and designs novel visualization and interaction methods that support exploration and understanding of the environments that this data represents.
\end{IEEEbiography}


\vfill


\end{document}